\documentclass[multphys,vecphys]{svmult}

\usepackage{makeidx}         
\usepackage{graphicx}        
\usepackage{multicol}        
\usepackage[bottom]{footmisc}

\makeindex             
\usepackage{amsmath}
\usepackage{amssymb}

\def\lea{\mathrel{\raise .4ex\hbox{\rlap{$<$}\lower 1.2ex\hbox{$\sim$}}}}
\def\gea{\mathrel{\raise .4ex\hbox{\rlap{$>$}\lower 1.2ex\hbox{$\sim$}}}}
\let\lsim=\lea

\begin{document}


\title*{Cosmic Microwave Background anisotropies: the power
spectrum and beyond}
\titlerunning{CMB anisotropies: the power spectrum and beyond}

\author{Enrique Mart\'{\i}nez-Gonz\'alez}
\institute{Instituto de F\'{\i}sica de Cantabria, CSIC-Universidad de Cantabria, Avenida Los Castros s/n, 39005 Santander, Spain
\texttt{martinez@ifca.unican.es}}
%
%
\maketitle

Most of the cosmological information extracted from the
CMB has been obtained through the power spectrum, however there is
much more to be learnt from the statistical distribution of the
temperature random field.  We review some recent developments
in the study of the Cosmic Microwave Background (CMB) anisotropies and
present a description of the novel tools developed to analyse the
properties of the CMB anisotropies beyond the power
spectrum.\footnote{This review was almost complete before 
the second release of WMAP data which was made publicly available 
on March 2006. 
The major improvement has been made in the polarization data which 
are now much better understood regarding both systematics and foregrounds. 
This has implied a significant change 
in the value of the optical depth ($\tau\approx 0.09$). Another difference with
respect to the values of the cosmological parameters estimated from the 
first year data is the possible 
significant deviation of the spectral index from unity as derived from 
the second release of data. Appart from these changes the rest of the 
parameters as well as the deviations from Gaussianity (``anomalies'') remain 
very much unaltered.}  

\section{Introduction}
\label{sec:introduction}
\sectionmark{Introduction}

The standard scenario of the universe includes an early inflationary
phase during which the universe experienced a drastic expansion. As a
consequence of this inflationary period the homogeneity, isotropy and
flatness of the universe that we observe today can be understood as a
natural outcome, without a fine tunning of the initial
conditions. Besides, quantum fluctuations of the fields dominating the
dynamics at that early phase constitute the seeds which, after
gravitational growing, will form the large scale structure (LSS) of
the universe at the present time (for a detailed description of
inflation and its consequences see e.g. \cite{liddle_lyth_00}).  The
statistical distribution of the quantum fluctuations of the field
responsible for inflation, known as inflaton, in the vacuum state is
Gaussian. In addition, the energy density fluctuations generated from
them are also Gaussian since they are related by linear
theory. Moreover, the anisotropies of the CMB are related to the
energy density fluctuations by the linearized Einstein-Boltzmann
equations, being therefore Gaussian distributed as well.  This is a
very important result since, for this particular statistical
distribution, all the properties of the temperature random field are
included in the second order moment. In harmonic space this is called
the radiation power spectrum, usually denoted by $C_{\ell}$.
 
The cosmological parameters which characterize the universe leave
their imprint in the radiation power spectrum. In the standard
inflationary scenario we only need to measure this second order moment
to obtain all the possible information carried out by the CMB. This is
in fact what many experiments have been doing with increasing
precision in recent years. The BOOMERANG \cite{de bernardis_00} and
MAXIMA \cite{hanany_00} experiments onboard stratospheric balloons were
among the first to establish the spatial flatness of the universe. Most
notably, the NASA Wilkinson Microwave Anisotropy Probe (WMAP) has
measured the $C_{\ell}$ with high precision up to the multipole
$\ell\approx 600$
\cite{bennett_03}. This measurement has implied a high improvement in
the determination of the cosmological parameters, reducing the
uncertainties below $10\%$ \cite{WMAP:parameters}. The uncertainties are
further reduced when other cosmological data sets, like the LSS distribution
of galaxies and SN Ia magnitude-redshift data, are combined with the
CMB anisotropies \cite{tegmark_04b}. Moreover, the degeneracy
among some parameters present when only a single data set is
considered is broken. This {\sl concordance model} has been further
confirmed by the first detection of polarization fluctuations by DASI
\cite{kovac_02} and the further determination of the polarization
power spectra by WMAP \cite{kogut_03}, DASI \cite{leitch_04}, CBI
\cite{readhead_04b} and BOOMERANG \cite{montroy_05,
piacentini_05}. An independent piece of evidence comes from the
cross-correlation between CMB anisotropies and fluctuations in the
galaxy number density.  Several authors have found positive
cross-correlations between the WMAP data and the NVSS radio galaxy
survey \cite{condon_98}, implying that the model of the universe is
different from Einstein-de Sitter and, in the case of a flat universe,
that the universe is dominated by a dark energy component
\cite{boughn-crittenden_04, fosalba_04, nolta_04, afshordi_04,
  vielva_06, mcewen_06a}.  

Measurements of the statistical distribution of the CMB temperature
and polarization random field represent a unique probe of the early
history of the universe. Thus, if the distribution is consistent with
Gaussianity at a high precision then it would imply a nice
confirmation of the standard inflationary model. On the contrary, if
some significant deviations from normality are found then this result
would contradict the standard single-field model and motivate the
search for alternative scenarios of the early history of the
universe. There have already been many suggestions of the latter in
the past based on many different physical phenomena. Most notably,
non-standard inflation models based on extensions of the standard one
in many different ways -e.g. multi-field scenarios as the curvaton
mechanism \cite{lyth_wands_02}, inhomogeneous reheating scenario, etc
(see \cite{bartolo_04} for a review)- and cosmic defects including global
and local strings, monopoles, domain walls and textures
\cite{vilenkin_shellard_94}. 

The structure of the paper is as follows. In Sec.~\ref{sec:theory} we
summarize the most relevant physical effects which generate the CMB
temperature and polarization
anisotropies. Sec.~\ref{sec:cosmological_constraints} deals with the
constraints imposed by recent CMB data sets on the cosmological parameters
as well as other cosmological data sets 
combined with the CMB. In Sec.~\ref{sec:beyond} we describe the
properties of the CMB anisotropies beyond the power spectrum and
discuss possible sources of non-Gaussianity, inhomogeneity and
anisotropy in their distribution on the celestial sphere. 
In Sec.~\ref{sec:methods} we discuss the 
methods developed for studies of the CMB temperature distribution and
of its global isotropy whereas in Sec.~\ref{sec:data_analysis} we review the analyses
performed so far on CMB data sets measured by different
experiments. Finally the conclusions are given in
Sec.~\ref{sec:conclusions}. 

\section{Theory}
\label{sec:theory}
\sectionmark{Theory}
The CMB anisotropies are produced by different physical effects acting
on this radiation before the last-scattering surface, when
the universe was around $380000$ years old. These {\sl primary
anisotropies}  
contain very valuable information on the early history of the
universe and can be observed today almost unaffected. After the
temperature of the photons drop below some $3000$ degrees they are
not able to ionize the hydrogen atoms and propagate freely. 
The physical effects producing the primary anisotropies at and
after recombination can be
summarised by the following equation \cite{martinez-gonzalez_90,
sanz_97}:
\begin{equation}  
{\frac {\Delta T} {T}} (\vec n ) \approx \frac {\phi_{e}(\vec n)}{3} +
2 \int_e^o \frac {\partial \phi}{\partial t} dt + {\vec n}
\cdot (\vec v_o -\vec v_{e}) + \Big( \frac {\Delta T}{T} (\vec
n)\Big)_{e}  
\label{eq:delta}
\end{equation}
The gravitational redshift suffered by the photons in their travel
from the last scattering surface to the observer is given by the first
two terms in the r.h.s. of eq.~\ref{eq:delta}. They are known as 
{\sl Sachs-Wolfe} and {\sl late Integrated Sachs-Wolfe} effect, respectively 
\cite{SW}. The
velocity of the baryon-photon fluid at recombination generates a
Doppler effect. The intrinsic temperature of the photons at
recombination, represented by the fourth term in the r.h.s. of
eq.~\ref{eq:delta}, also contributes to the total
anisotropy. Eq.~\ref{eq:delta} accounts for the anisotropies at
recombination. Before recombination, changes in the
gravitational potential, due to the imperfect baryon-photon coupling, can also
produce a gravitational redshift (the {\sl early Integrated Sachs-Wolfe effect}). The
accurate 
computation of all the contributions requires to solve the linearized coupled
Einstein-Boltzmann equations. Several codes have been developed to
numerically solve those equations as a function of the cosmological
parameters (about 10, see below), being some of them publicly
available (e.g. CMBFAST 
\cite{CMBFAST}, CAMB \cite{lewis_00}). A
comparison among the results of different codes showed that the
accuracy achieved is very good, reaching $\approx 0.1\%$ up to $\ell =3000$
\cite{seljak_03}.      

\begin{figure}
\centering
\includegraphics[height=3.in]{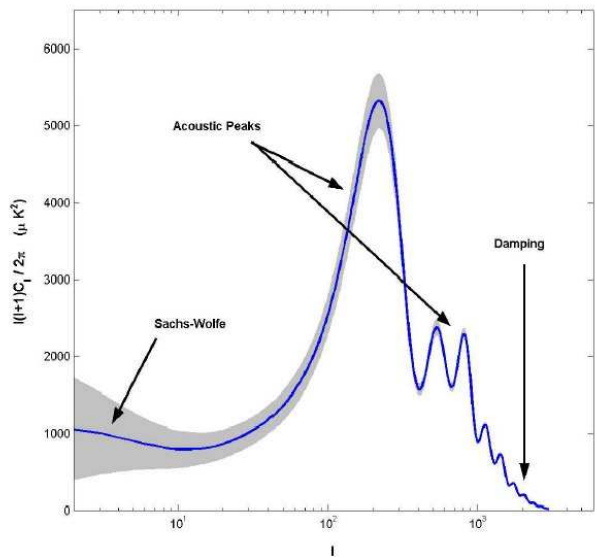}
\caption{CMB power spectrum $C_\ell$ for the best fit model given in 
\cite{bennett_03}. The gray band represents the cosmic variance. The
spectrum has been computed with the CMBFAST code \cite{CMBFAST}.}
\label{fig:cl}       
\end{figure}

As will be explained below, a very relevant statistical quantity to compute
is the 2-point correlation function of the temperature anisotropies or
equivalently  the correlation of the spherical harmonic coefficients:
\begin{equation}
<a_{\ell m}a_{\ell' m'}^*> = C_\ell \delta_{\ell\ell'}\delta_{mm'},
\label{eq:cl}
\end{equation}
where $C_\ell$ is the anisotropy power spectrum and the $a_{lm}$ are
the coefficients of the spherical harmonic expansion
\begin{equation}
\frac {\Delta T}{T} (\vec n)= \sum_{\ell,m} a_{\ell m} Y_{\ell m}(\vec n),
\label{eq:expansion}
\end{equation}
The homogeneity and isotropy of the universe have been assumed in
eq.~\ref{eq:cl}, 
i.e. the $a_{lm}$ are not correlated for different $l$ or $m$.  
In Fig.~\ref{fig:cl} the scales at which the different effects dominate 
the power spectrum $C_\ell$ are marked. As we can see, the
gravitational effects dominate at the lower multipoles (large angular
scales). At intermediate scales, $100<\ell<1000$, the spectrum is
dominated by several oscillations, usually called {\sl acoustic peaks}.  
These peaks appear as a consequence of the balance between the
gravitational force and the radiation pressure. At the smallest scales,
$\ell>1000$, the $C_\ell$ are damped because of the width of
recombination and the imperfections in the coupling of the
photon-baryon fluid ({\sl Silk effect}, \cite{silk_68}). 

When the small scales in the matter distribution become nonlinear
and their collapse give rise to the formation of the first stars and
quasars they start to reionize the surrounding matter. This ionized
matter can interact again with the microwave photons and produce new
anisotropies. {\sl Secondary anisotropies} can be either produced by
the scattering with free electrons or by the gravitational effect of
the matter density evolution. The scattering can leave a very clear
imprint when the photons happen to cross the core of rich galaxy
clusters where the electron density and temperature are very high (of
several $10^{-3}$cm$^{-3}$ and $\sim 10^8$K, respectively). In this case the
electrons inject energy to the photons through inverse Compton
scattering, producing a distorsion of the black-body spectrum. This
effect is called {\sl the Sunyaev-Zeldovich} effect
\cite{SZ}.\\
The evolving gravitational wells in the large scale structure produce
a gravitational redshift in the photons. This is known as the {\sl
late Integrated Sachs-Wolfe} effect \cite{SW} or {\sl the Rees-Sciama}
effect when the evolution is non-linear \cite{RS}. Besides, the
trajectory of the photons is lensed by the same gravitational wells
producing a noticeble effect on the CMB anisotropies at arcmin angular
scales. One important issue in the observation of the CMB anisotropies
is therefore to distinguish between primary and secondary
anisotropies. In the case of the SZ effect, as the frequency
dependence is different from the Planckian one, it is possible to
separate the effect from the intrincsic CMB anisotropies by using
multifrequency observations. However, for the lensing effect this is
not possible and other properties, such as high-order correlations in
the CMB temperature and polarization fields,
have to be used (see e.g. \cite{hu_00,
goldberg_spergel_99, seljak_zaldarriaga_99}).   

\begin{figure}
\centering
\includegraphics[height=4.5in, angle=270]{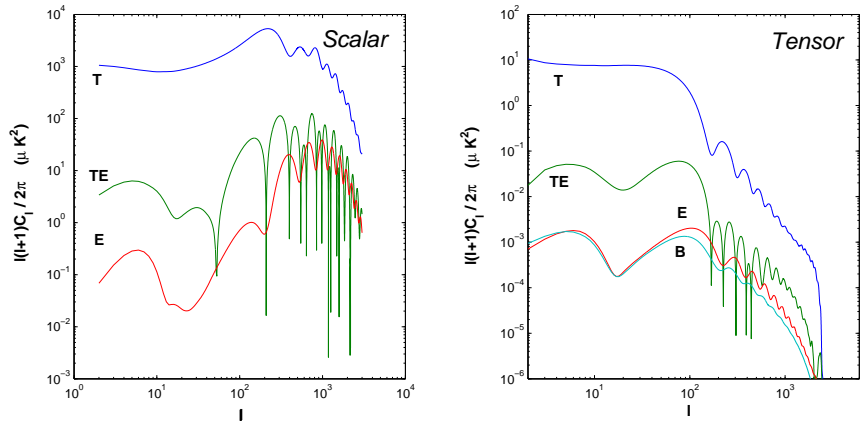}
\caption{Angular power spectra for temperature and polarization. The
$C_\ell$'s are plotted for scalar perturbations on the left and for
tensor perturbations on the right, where a tensor/scalar ratio of
$r=0.01$ has been chosen. The rest of the parameters have been fixed
to the best fit model of \cite{bennett_03}. All the spectra have been
computed with the CMBFAST code.}
\label{fig:spectra_model}       
\end{figure}

In the same manner that anisotropies in the temperature are expected
in the gravitational instability scenario for structure formation,
anisotropies in the polarization of the CMB are also expected. The
physical effects giving rise to these anisotropies are, however,
different. Linear polarization is generated by Thompson scattering at
the end of recombination when the growth of the mean free path of the
photons allows temperature anisotropies to grow. In particular a
quadrupole is formed in the reference frame of each electron producing
polarization after the scattering (see, e.g.,
\cite{challinor_05}). The expected level of polarization is however
small $\approx 5\%$. As shown by \cite{zaldarriaga_seljak_97,
kamionkowski_97} two rotationally invariant quantities $E, B$ can be
constructed from the Stokes parameters $Q, U$. The different behaviour
of $E, B$ under parity transformations implies that only three spectra
are needed to characterise CMB polarization: $C_\ell^E, C_\ell^B,
C_\ell^{TE}$.

In addition to the anisotropies generated by physical effects
associated to the energy-matter density perturbations ({\sl scalar
perturbations}), new anisotropies can be generated by a background of
primordial gravitational waves ({\sl tensor perturbations}). This
background is also a generic prediction of inflation. The primordial
power spectra for both scalar and tensor perturbations are usually
characterised by a scale-free law of the form:
\begin{equation}  
P_s(k) =A_s (k/k_0)^{n_s}, P_t(k) =A_t (k/k_0)^{n_t}, 
k_0=0.05Mpc^{-1}. 
\end{equation}
There is no observational evidence of the primordial background of
gravitational waves yet. Upper limits have been imposed by the
combination of WMAP and other high resolution CMB data with large
scale structure data implying
a tensor-to-scalar amplitude ratio $r\equiv A_t/A_s<0.9$
\cite{bennett_03}, see next sections. In any case,
the amplitude of both temperature and polarization power spectra
produced by tensor perturbations are believed to be several orders of
magnitude below the ones corresponding to the scalar
perturbations. However, the $B$-mode can only be generated by
gravitational waves and thus its detection represents a unique
evidence of the existence of this primordial background. This is the
reason why there is now so much interest in planning new more
sensitive CMB experiments to detect polarization (see
e.g. \cite{bouchet_05}). The detection of the $B$-mode is
expected to be much harder than the already detected $E$-mode because
of its intrinsically smaller amplitude and the relatively larger foreground
emissions (expected from Galactic synchroton and thermal
dust and extragalactic radio sources; for a recent discussion of these isues see
\cite{tucci_05}). An additional complication comes from the lensing
conversion of the $E$-mode to the $B$-mode
\cite{zaldarriaga_seljak_98} which is expected to
dominate over the primordial $B$-mode at multipoles $\ell \gtrsim 100$
\cite{knox_02}. Several methods have been developed to reconstruct the
gravitational lensing potential from the E and B-mode polarization
correlations and remove the lensing contamination \cite{hu_okamoto_02,
kesden_03, seljak_hirata_04}. A more technical problem is the separation
of the mixing of $E$ and $B$-modes for observations with partial sky
coverage for which other methods have been proposed \cite{lewis_02,
bunn_03, challinor_chon_05}.    

Outside the inflationary scenario there are other
possibilities to generate $B$-modes. For instance, cosmic strings or
primordial magnetic fields would generate a vector and a vector plus
tensor components of metric
perturbations, respectively, contributing to the $B$-mode polarization
(see e.g. \cite{pogosian_03, lewis_04}). It is important to remark,
however, that in the case of cosmic strings and other topological
defects their role as seeds for the large scale
structure formation is already very much constrained by the observed
$C_\ell$,  
see below.  

In Fig.~\ref{fig:spectra_model} the different temperature and
polarization power spectra corresponding to the primary anisotropies
for the concordance model are
plotted for the scalar and tensor perturbations assuming a value of $r=0.01$.  
The power of the tensor perturbations is a few orders of magnitude below the
corresponding scalar power spectra for $\ell \lesssim 100$ and decays
strongly for larger multipoles.
 
\begin{figure}
\centering
\includegraphics[height=4.5in, angle=270]{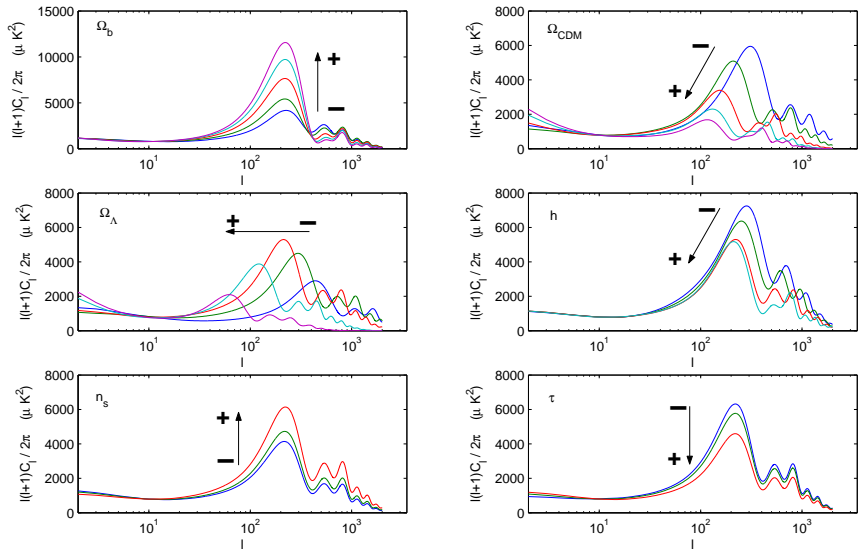}
\caption{Dependence of the temperature power spectrum
$C_\ell$ on some relevant cosmological parameters (the spectra have been
computed with the CMBFAST code).}
\label{fig:parameter_dependence}       
\end{figure}

Many parameters are needed to characterize the standar cosmological
model. Their variation changes the amplitude and shape of the
temperature and polarization power spectra in many different ways (see
Fig.~\ref{fig:parameter_dependence} for their effect on the
temperature power spectrum). The cosmological parameters can be
classified depending on whether they characterize the background
universe or the primordial power spectrum.\\   
The background
Friedmann-Robertson-Walker universe and its matter and energy content
are determined by the following parameters: {\sl physical baryonic
density}, $w_b 
= \Omega_b h^2$; {\sl physical matter density}, $w_m = \Omega_m h^2$,
where $\Omega_m$ is the matter density parameter including the
contributions from baryons, cold dark matter and neutrinos, $\Omega_m
=\Omega_b + \Omega_{CDM} + \Omega_\nu$; {\sl physical neutrino
density}, $w_\nu = \Omega_\nu h^2$; {\sl dark energy equation of state
parameter}, $w\equiv p_{DE}/\rho_{DE}$; {\sl dark energy density},
$\Omega_{DE}$ (in case $w=-1$ the dark energy takes the form of a
cosmological constant and its energy contribution is denoted by
$\Omega_{\Lambda}$) and {\sl the Hubble constant}, $h\equiv H_0/100$km
s$^{-1}$ Mpc$^{-1}$.\\ 
The scalar and tensor primordial power spectra
are characterized by the following parameters: {\sl amplitude of the primordial
scalar power spectrum}, $A_s$, defined by $P_s(k) =A_s (k/k_0)^{n_s}$,
where $k_0=0.05$Mpc$^{-1}$; {\sl scalar spectral index}, $n_s$; {\sl
running index}, $\alpha=dn_s/d\ln k$, normally determined at the scale
$k_0=0.05$Mpc$^{-1}$; {\sl tensor-to-scalar ratio}, $r=A_t/A_s$; {\sl
tensor spectral index}, $n_t$, which is normally assumed to be
$n_t=-r/8$ from the consistency relation of inflation (see
e.g. \cite{liddle_lyth_00}).\\  
There is an extra
parameter that accounts for the reionization history of the universe:
{\sl the optical depth}, $\tau = \sigma_T
\int^{t_0}_{t_r} n_e(t) dt$, where $\sigma_T$ is the Thompson
cross-section and $n_e(t)$ is the electron number density as a
function of time.\\
Besides, there are two possible types of matter density
fluctuations: {\sl adiabatic} which preserve the entropy per
particle and {\sl isocurvature} which preserve the total energy density. 
The standard inflationary scenario predicts fluctuations of the
adiabatic type.\\
It is important to
note that the manner in which some of the parameters enter in the
calculation of the $C_\ell$ produce degeneracies. In particular, there
is the well known geometrical degeneracy involving $\Omega_m$,
$\Omega_{\Lambda}$ and $\Omega_k$, where $\Omega_k$ is the curvature
density parameter, $\Omega_k \equiv 1 - \Omega_m - \Omega_{\Lambda}$
(an example of this degeneracy can be found in Fig.~5 of
\cite{martinez-gonzalez_vielva_05}). This fact makes clear the need to
combine different cosmological data sets to break those degeneracies,
as will be shown below.    

\section{Cosmological constraints}
\label{sec:cosmological_constraints}
\sectionmark{Cosmological constraints}

Several issues are important to bear in mind in the analysis of the
CMB anisotropy data. First of all, the microwave sky is interpreted as a
realization of the temperature anisotropy random field whose origin is
in the inflationary era during the first moments in the history of the
universe. The quantum fluctuations in the field dominating the
dynamics at that epoch, called the inflaton field, generate
energy perturbations which produce anisotropies in the CMB through the physical
effects discussed in the previous section. The anisotropy field so
generated is not ergodic on the sphere and so averages on the field
do not coincide with averages on the sphere. This fact implies an
intrinsic uncertainty in the statistical quantities extracted from the
data, in particular in the power spectrum $C_\ell$, called {\sl the
cosmic variance}. For the standard model this uncertainty is easy to
compute since the random field is expected to be Gaussian, $\Delta
C_\ell = C_\ell/\sqrt{\ell+0.5}$ (this is valid for the whole sky, however
when only a fraction of the sky $f_{sky}$ is observed then the error
is increased by approximately $f_{sky}^{-1/2}$). The value of this
uncertainty in plotted in Fig.~\ref{fig:cl} for the standard model.\\
Second, at the millimeter wavelenghts where the intensity of the
black-body spectrum with temperature $\approx 2.725$K is maximum there
are other astronomical {\sl foregrounds} which are also bright. 
These are the Galactic foregrounds: synchrotron,
free-free and thermal dust, and also the extragalactic sources whose
emission peaks in the radio or the infrared bands. The problem of
separating the foreground emission from the cosmic signal is crucial
and requires multifrequency observations and a good understanding of
the frequency and spatial properties of the foregrounds. In recent
years a strong effort in this direction is being made by the CMB
community, devising new instruments optimized to observe at
frequencies around $100$GHz where the CMB emission dominates over the
foregrounds (for both temperature and polarization, see
e.g. \cite{tucci_05}), and developing sophisticated
methods of component separation (see e.g. \cite{delabrouille_06,
barreiro_06a} in this volume). In any case, there are still regions 
of the sky (near the Galactic plane and others dominated by
extragalactic sources) where the CMB is unavoidably hampered by the
foregrounds. These contaminated regions are usually masked and therefore
removed from the data analyses. 

In the next subsections we will discuss the results obtained on the
model of the universe based on the $C_\ell$. Following the standard
model and some observational evidences, the anisotropies are assumed
to be Gaussian and thus all the statistical information is included
in the second-order moments. It is for this reason that most of the
analyses performed to constrain the cosmological parameters are based
on the $C_\ell$. Even the likelihood methods to derive those
parameters are based on the Gaussian assumption (see for instance the
review by \cite{hu_dodelson_02}). In
Sec.~\ref{sec:methods} and after we will also review the works
made to test the Gaussian assumption.         

\subsection{Cosmological parameters from the $C_\ell$}
\label{subsec:parameters}

In the last decade there has been a strong technological development
in CMB devices allowing very sensitive measurements of the temperature
of the microwave sky. Due to the atmospheric emission at microwave
frequencies the experiments have to be located at very dry places on
Earth or onboard of balloons or satellites. In particular the space
mission NASA WMAP has made 
possible to cover the whole sky from a priviledged position (the
Lagrangian point 2 of the Sun-Earth system), saved from contaminating 
emissions coming from the
Earth, Moon and Sun. In table~\ref{tab:WMAP} the values of
the cosmological parameters determined with the first year data of
that mission are shown. An 
interesting result that can be derived from this table is that the
Einstein-de-Sitter model (i.e. flat geometry with null dark energy
density) is many standard deviations away from the best fit 
model (asuming a flat universe and an optical depth $\tau <0.3$)
\cite{WMAP:parameters}.   

\begin{table}
\centering
\caption{Cosmological parameters using only WMAP first year
data. In the fit the universe is assumed to be spatially flat and the
value of the optical depth is constrained to $\tau<0.3$ (from
\cite{WMAP:parameters}).  The physical baryonic density parameter
$w_b$ is defined 
by $w_b=\Omega_b h^2$ and similarly for the physical matter density
$w_m$ which includes the contribution from all the matter species:
baryons, cold dark matter (CDM) and neutrinos.}
\label{tab:WMAP}       
\begin{tabular}{cc}
\hline\noalign{\smallskip}
\it Parameter& \it Values (68\% CL)\\
\noalign{\smallskip}\hline\noalign{\smallskip}
$w_b$ & 0.024$\pm$0.001\\
$w_m$ & 0.14$\pm$0.02\\
$h$ & 0.72$\pm$0.05\\
$A_s$ & 0.9$\pm$0.1\\
$\tau$ & 0.166$_{-0.071}^{+0.076}$\\
$n_s$ & 0.99$\pm$0.04\\
\noalign{\smallskip}\hline
\end{tabular}
\end{table}

\begin{figure}
\centering
\includegraphics[height=4.5in, angle=270]{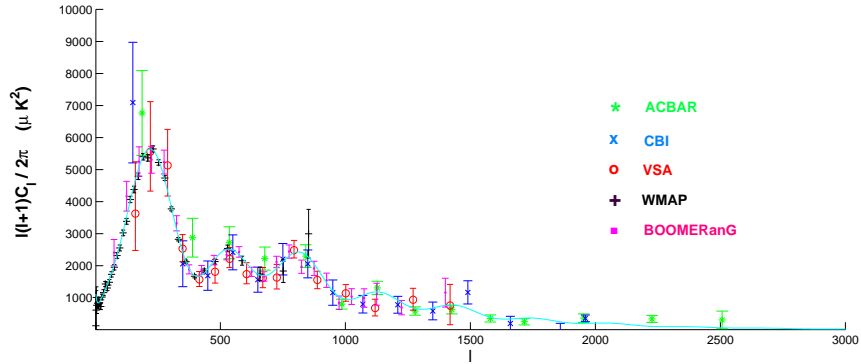}
\caption{The temperature power spectrum $C_\ell$ measured by WMAP, ACBAR,
CBI, VSA and BOOMERANG, compared to the best fit model given by
\cite{bennett_03}.} 
\label{fig:temperature_model}       
\end{figure}

The resolution of the WMAP experiment allows a good precision in the
determination of the $C_\ell$ up to multipoles $\ell \lesssim
600$. Other experiments based on the ground like ACBAR \cite{kuo_04}, CBI
\cite{readhead_04a}, VSA \cite{dickinson_04} and
balloon-borne like BOOMERANG \cite{jones_05} are able to
reach higher resolutions covering small patches on the sky with good
sensitivity. The data obtained 	
with these experiments therefore complement very well the WMAP ones.
In Fig.~\ref{fig:temperature_model} the temperature anisotropy results
from WMAP and the other high resolution experiments as a function of
the multipole are shown. We can see that the concordance model follows
quite well the data up to $\ell\approx 2000$. For higher multipoles only ACBAR
and CBI experiments have enough resolution and the data seem to
indicate an excess of power as compared to the model prediction. This
excess can be interpreted as a contribution from secondary anisotropies
coming from the SZ effect \cite{bond_05} or/and the emission of radio sources 
\cite{toffolatti_05}. However, new multifrequency data at arcminutes
resolution are necessary to
confirm the excess and discriminate among possible causes.

\begin{figure}
\centering
\includegraphics[height=4.5in, angle=270]{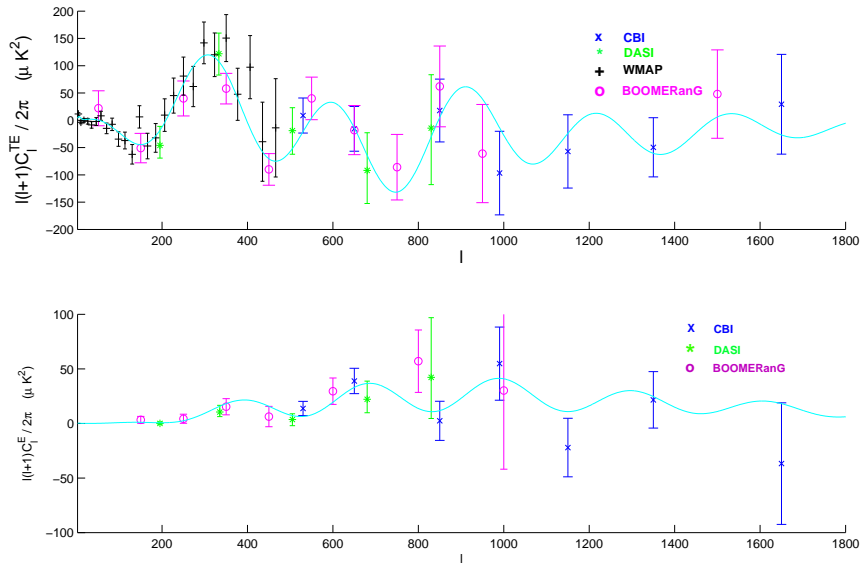}
\caption{Polarization power spectra measured by DASI, WMAP, CBI and
BOOBMERANG (TE) and DASI, CBI and BOOMERANG (E). Also plotted is the best fit
model given by \cite{bennett_03}.} 
\label{fig:polarization_model}       
\end{figure}

Since the recent detection of polarization by the DASI experiment
\cite{kovac_02} several experiments have measured the $TE$ cross-power
spectrum $C_\ell^{TE}$ (WMAP \cite{kogut_03}, CBI \cite{readhead_04b}
and BOOMERANG \cite{piacentini_05}) and 
the $EE$ one $C_\ell^E$ (CBI \cite{readhead_04b} and BOOMERANG
\cite{montroy_05}). The sensitivity of the polarization data is not yet
comparable to the temperature one. However, it is already sufficient
to confirm the main features of the concordance model and, more
specifically, the adiabatic nature of the primordial matter density
fluctuations. Namely, that the peaks of the polarization power
spectrum, $C_\ell^E$, are out of phase with the temperature
$C_\ell$ ones. The $TE$ cross-power spectrum, $C_\ell^{TE}$, and the
$E$-mode polarization power spectrum, $C_\ell^E$, are shown in Fig.~\ref{fig:polarization_model}. 

The next challenge is the detection of the $B$-mode polarization. As
commented in the previous section this detection would
unambiguosly indicate the existence of a background of primordial
gravitational waves. Moreover, it is the only chance that we have to
know about its possible existence in the next decade since experiments aimed to
directly detect gravitational waves are still not sensitive
enough. The amplitude of this background, $A_t$, or equivalently the
tensor-to-scalar ratio $r$, is proportional to the energy scale of
inflation. Whereas the all-sky ESA Planck space
mission\footnote{http://www.rssd.esa.int/Planck}      
\cite{planck} will be limitted by instrument 
sensitivity, being able to detect values down to $r\approx 0.05$ \cite{tucci_05},
planned very sensitive ground-based experiments covering small patches
of the sky and carrying arrays of
1000's of detectors like PolarBear \cite{polarbear},
Clover\cite{clover}, BRAIN \cite{brain} or QUIET \cite{quiet}, are 
expected to be limitted by the ability to remove the foregrounds and
are expected to reach values of $r\approx 0.01$ in the best case. In
order to significantly improve this limit we will have to wait for the
next generation of space missions with $10^3-10^4$ detectors, now under
discussion by ESA, NASA and other space agencies \cite{white_05,
bouchet_05}. The combination of complete sky coverage, 
very sensitive multifrequency observations and an optimal control of
systematics make the space missions the ultimate experiments to go
down to values $r\sim 10^{-3}-10^{-4}$ allowing to probe energy scales
for inflation down to $\approx 5\times 10^{15}$GeV \cite{tucci_05}.

\subsection{Combination with other cosmological data sets}
\label{subsec:combination}

\begin{figure}
\centering
\includegraphics[height=4.5in]{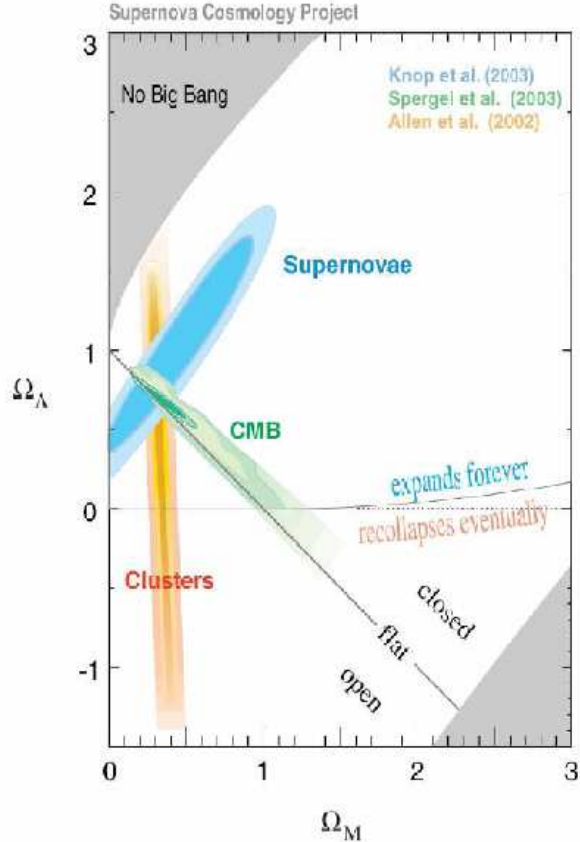}
\caption{Confidence contours for the pair ($\Omega_m, \Omega_\Lambda$)
combining SN Ia, CMB and cluster density data (taken from the
Supernova Cosmology Project).}
\label{fig:SN1}       
\end{figure}

The first observations showing the existence of a dark energy
dominating the dynamics of the universe were those based on the
{\sl luminosity distance-redshift diagram} determined with supernovae SN
Ia \cite{riess_98, perlmutter_99}. Those results were, however, taken with certain caution because of
the assumption made that the low and high redshift SN Ia had the same light
curve behaviour (in addition, there were many other possible
systematics which raised some concern). In any case, SN Ia data
complements very 
well the CMB anisotropies since for the former the dependence of the
model prediction 
on the dark matter and dark energy density parameters enters approximately as
$\Omega_m - \Omega_\Lambda$, and as $\Omega_m + \Omega_\Lambda$ for the
latter (see Fig.~\ref{fig:SN1}). 

The large galaxy surveys, {\sl 2-degree Field Galaxy Redshift Survey} 
2dFGRS \cite{percival_01} and {\sl Sloan Digital Sky Survey} SDSS
\cite{tegmark_04a} have recently 
reached $\approx 200000$ measured redshifts on large fractions of the
sky. A three-dimensional power spectrum $P(k)$ of density
perturbations is estimated with each data set. The cosmological
parameters enter in the prediction of $P(k)$ through the initial power
spectrum ($A_s, n_s$) and the transfer function which linearly
connects the initial and present spectra. Besides, an additional
parameter, {\sl the bias $b$}, is required to link the galaxy power spectrum
to the matter one $P(k)$. Although this parameter will in principle
depend on the scale, however, it is found that at large scales $b$ is
scale-independent (see e.g. \cite{tegmark_04b}). There are still other
problematics 
that need to be corrected before using the data for accurate parameter
determination such as the redshift-space distorsions due to galaxy
peculiar velocities, survey geometry effects, ... Once those effects
are corrected the galaxy redshift surveys play a key role
in breaking the CMB degeneracies. For instance, the degeneracies of
the model predictions on 
both data sets on the plane formed by the 
spectral shape parameter $h\Omega_m$ and the baryon fraction
$\Omega_b/\Omega_m$ are almost orthogonal.

\begin{table}
\centering
\caption{Cosmological parameters from WMAP,
CBI, ACBAR and 2dFGRS combined data (from \cite{bennett_03}).}
\label{tab:WMAP_LSS}       
\begin{tabular}{cc}
\hline\noalign{\smallskip}
\it Parameter& \it Values (68\% CL)\\
\noalign{\smallskip}\hline\noalign{\smallskip}
$w_b$ & 0.0224$\pm$0.0009\\
$w_m$ & 0.135$_{-0.009}^{+0.008}$\\
$w_\nu$ & $<$ 0.0076 (95\% CL)\\
$w$ & $<$ -0.78 (95\% CL)\\
$\Omega_{DE}$ & 0.73$\pm$0.04\\
$h$ & 0.71$_{-0.03}^{+0.04}$\\
$\tau$ & 0.17$\pm$0.04\\
$A_s$ & 0.833$_{-0.083}^{+0.086}$\\
$n_s$ & 0.93$\pm$0.03\\
$\alpha$ & -0.031$_{-0.018}^{+0.016}$\\
$r$ & $<$ 0.90 (95\% CL)\\
\noalign{\smallskip}\hline
\end{tabular}
\end{table}

\begin{figure}
\centering
\includegraphics[height=5.in]{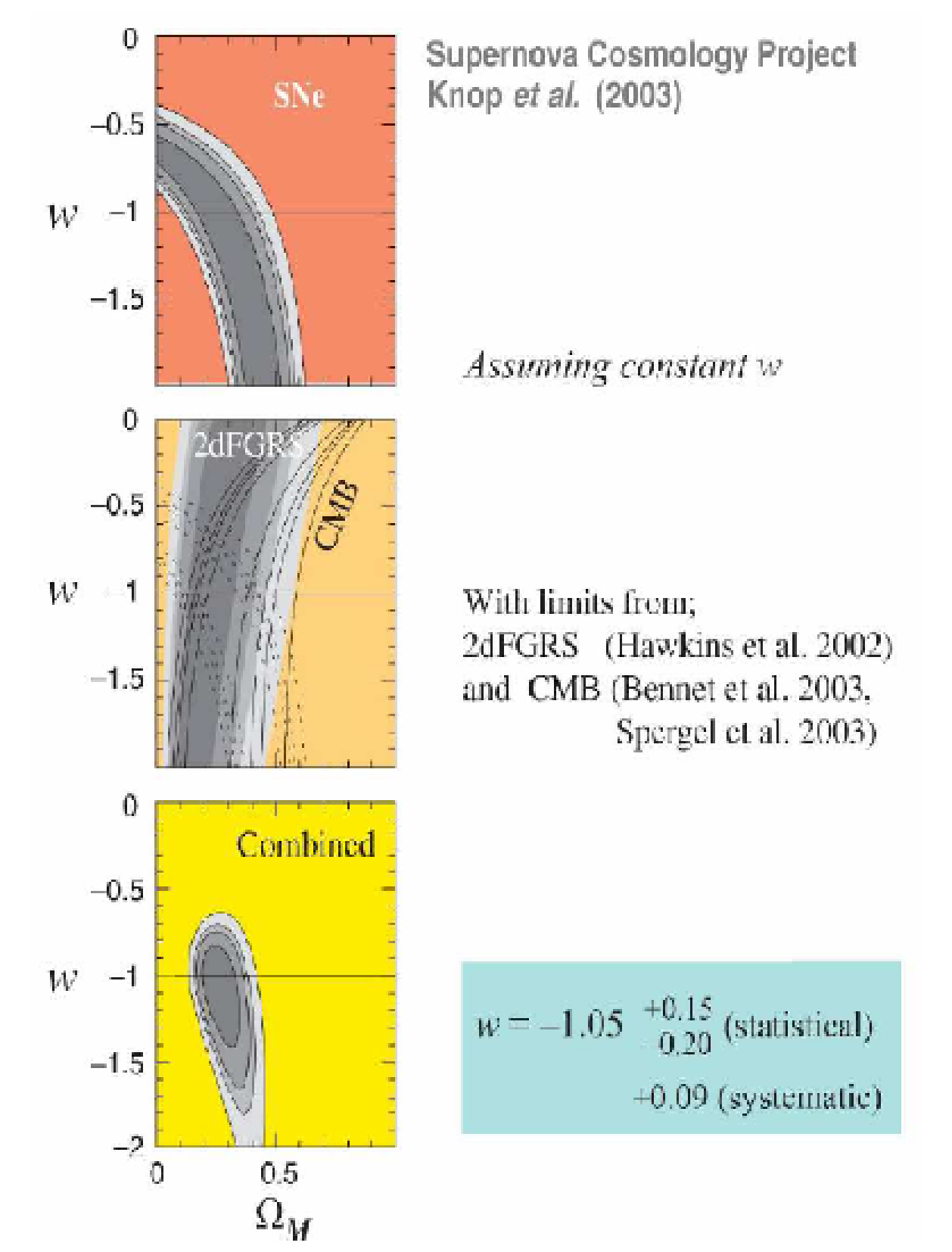}
\caption{Confidence contours for the pair ($\Omega_m, w$) using either
SN Ia, 2dFGRS and CMB or a combination of the three data sets (taken from the
Supernova Cosmology Project).}
\label{fig:SN2}       
\end{figure}

Results on 11 free cosmological parameters combining CMB data (WMAP
\cite{hinshaw_03}, CBI \cite{readhead_04a}, ACBAR 
\cite{kuo_04}) with the 2dFGRS galaxy redshift survey
\cite{percival_01} are given in table~\ref{tab:WMAP_LSS}. As can be
seen from that table, the combination of different cosmological data
sets has allowed for the first time an accuracy $\lesssim 10\%$ in the
determination of most parameters. Results combining WMAP data
with the SDSS galaxy redshift survey \cite{tegmark_04a} have also
produced similar values for the parameters \cite{tegmark_04b}.\\
In addition to CMB and galaxy surveys, other combinations including 
the HST key project value for the Hubble parameter \cite{freedman_01},
SN Ia magnitude-redshift 
data \cite{riess_01, tonry_03}, Ly$\alpha$ forest power spectrum
\cite{croft_02, mcdonald_04} or abundancies of reach clusters of
galaxies can help to improve the results
\cite{WMAP:parameters, seljak_05, rapetti_05, jassal_05}. Confidence
contours in the plane ($\Omega_m, \Omega_\Lambda$) for the combination
CMB+SN Ia+cluster abundancies are given in Fig.~\ref{fig:SN1}. The
complementarity of the three data sets to break degeneracies is
clearly shown. A similar effect in the plane ($\Omega_m, w$) is
obtained by combining CMB+2dFGRS+SN Ia (see Fig.~\ref{fig:SN2}).  

\begin{figure}
\centering
\includegraphics[height=3.5in, angle=90]{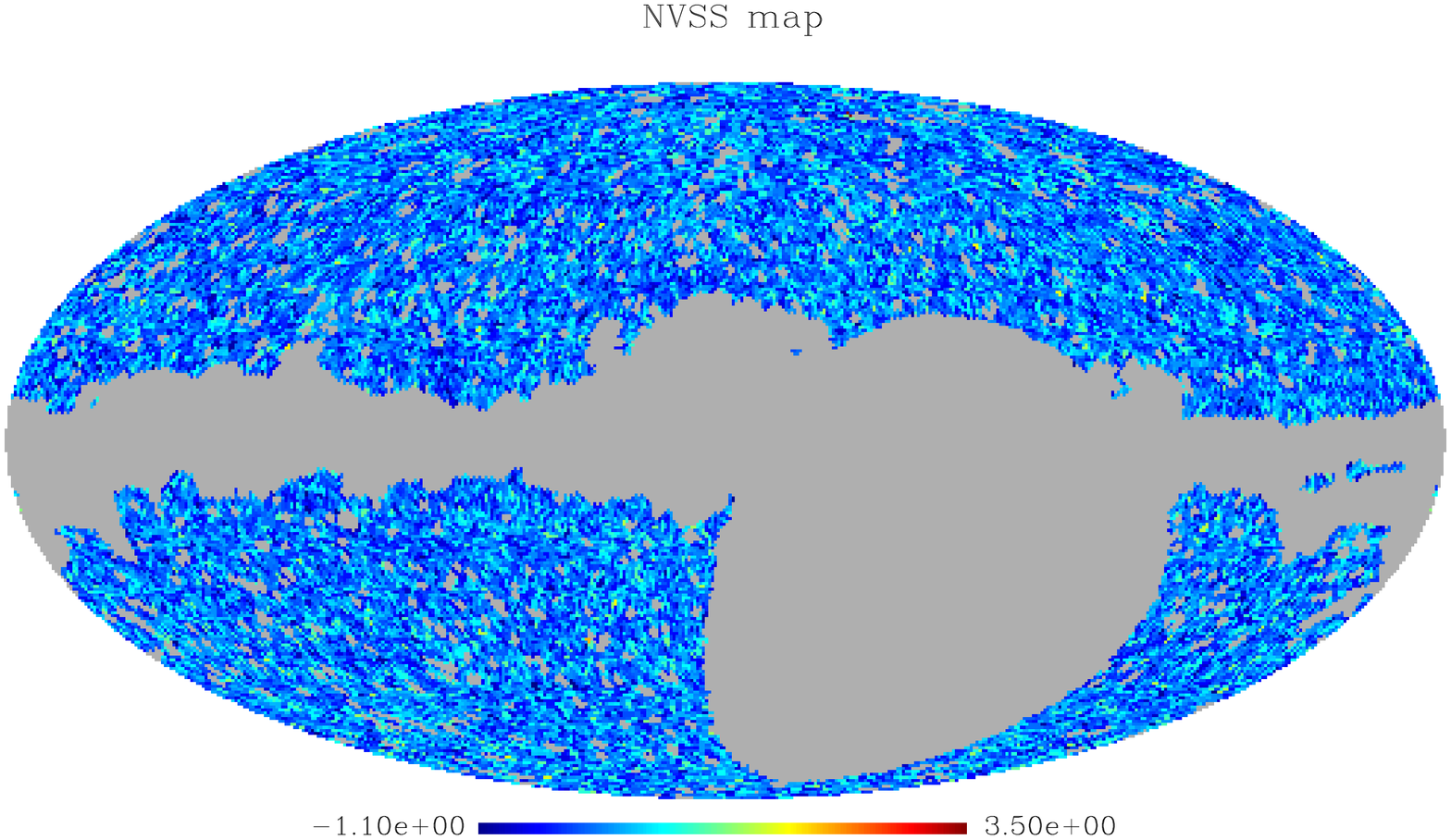}
\includegraphics[height=3.5in, angle=90]{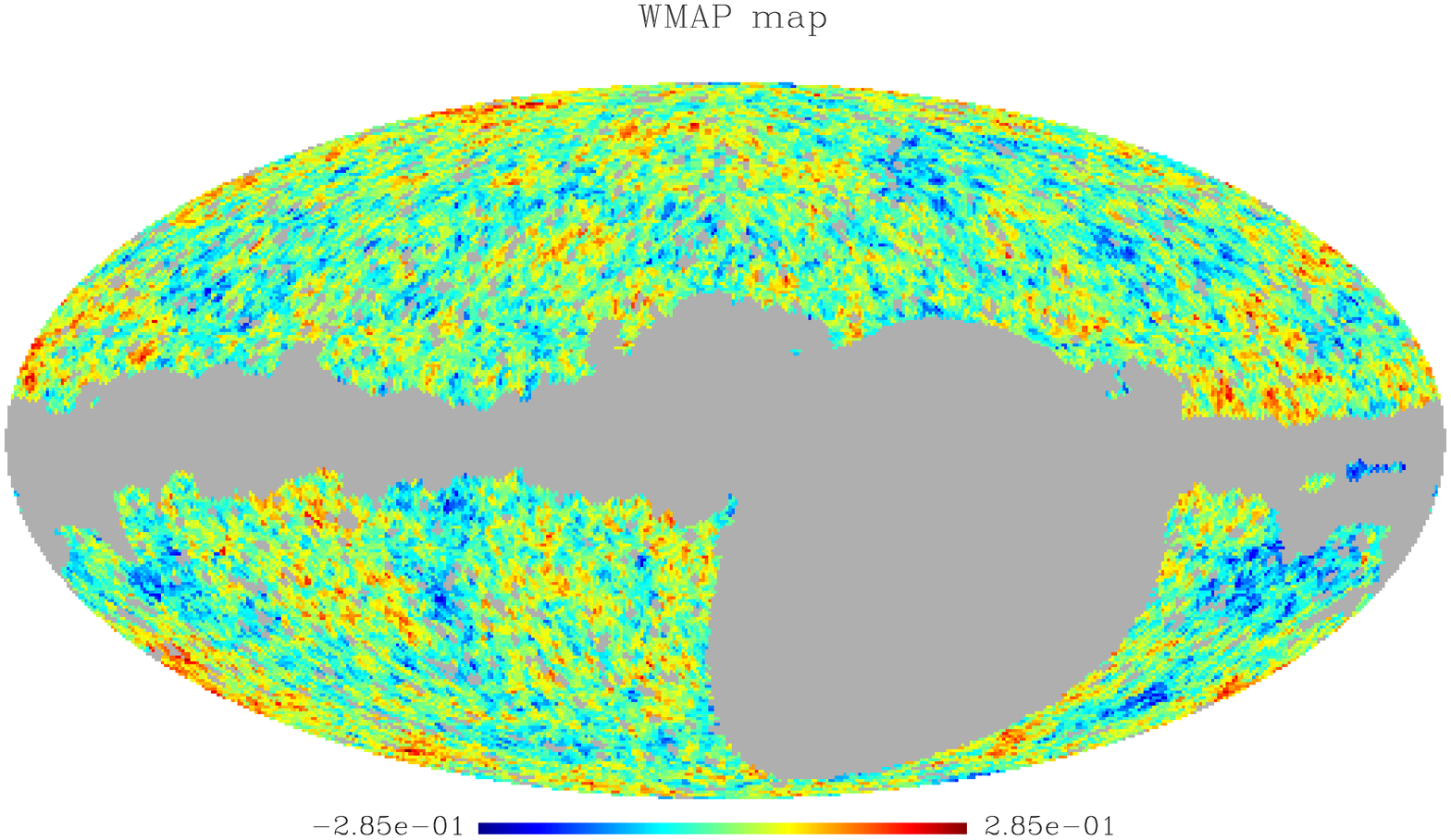}
\caption{The radio sources NVSS and the combined WMAP maps including
the common masks used in the cross-correlation analyses (taken from
\cite{vielva_06}).}  
\label{fig:nvss_wmap}       
\end{figure}

\begin{figure*}
\centering
\includegraphics[height=2.5in]{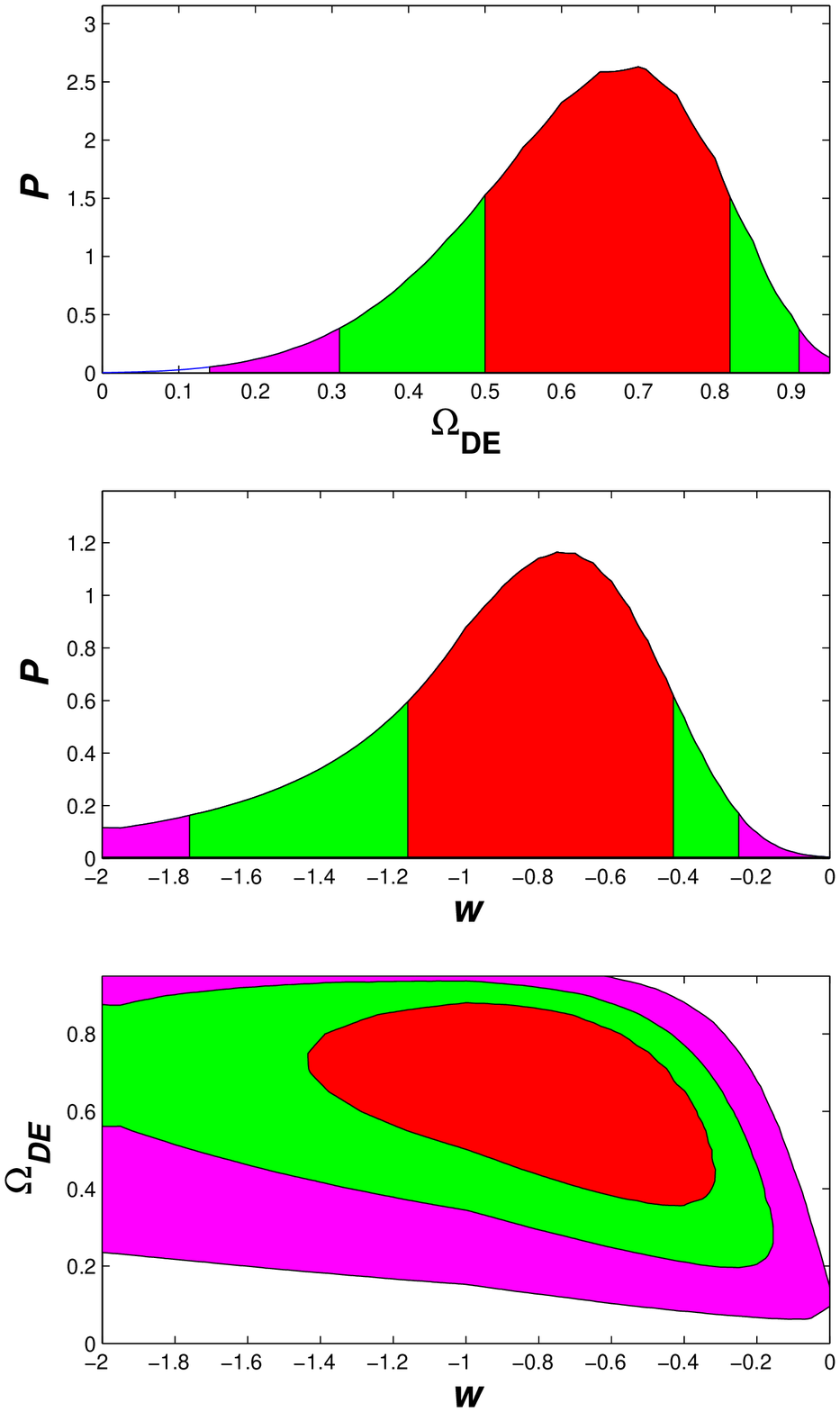}
\includegraphics[height=2.5in]{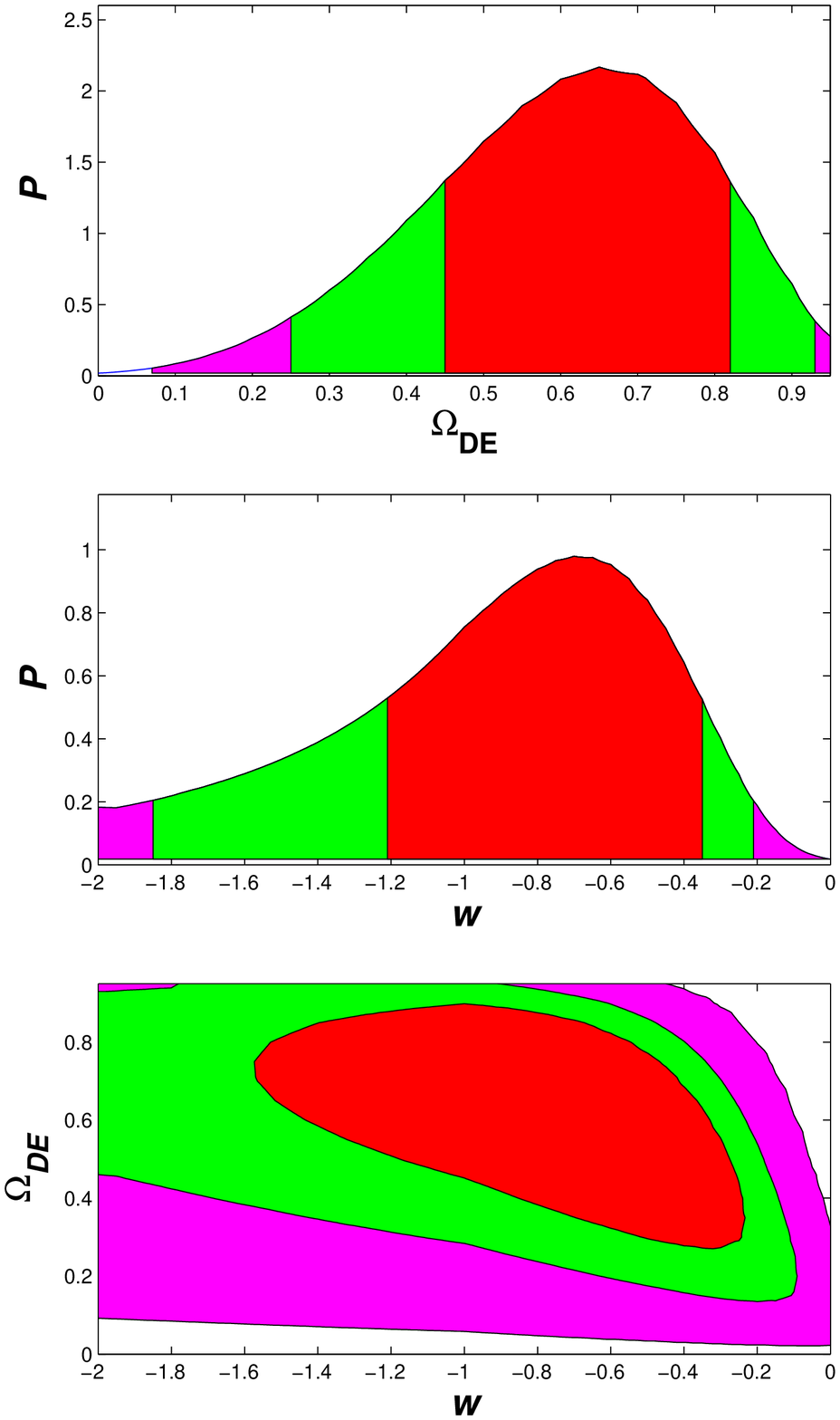}
\includegraphics[height=2.5in]{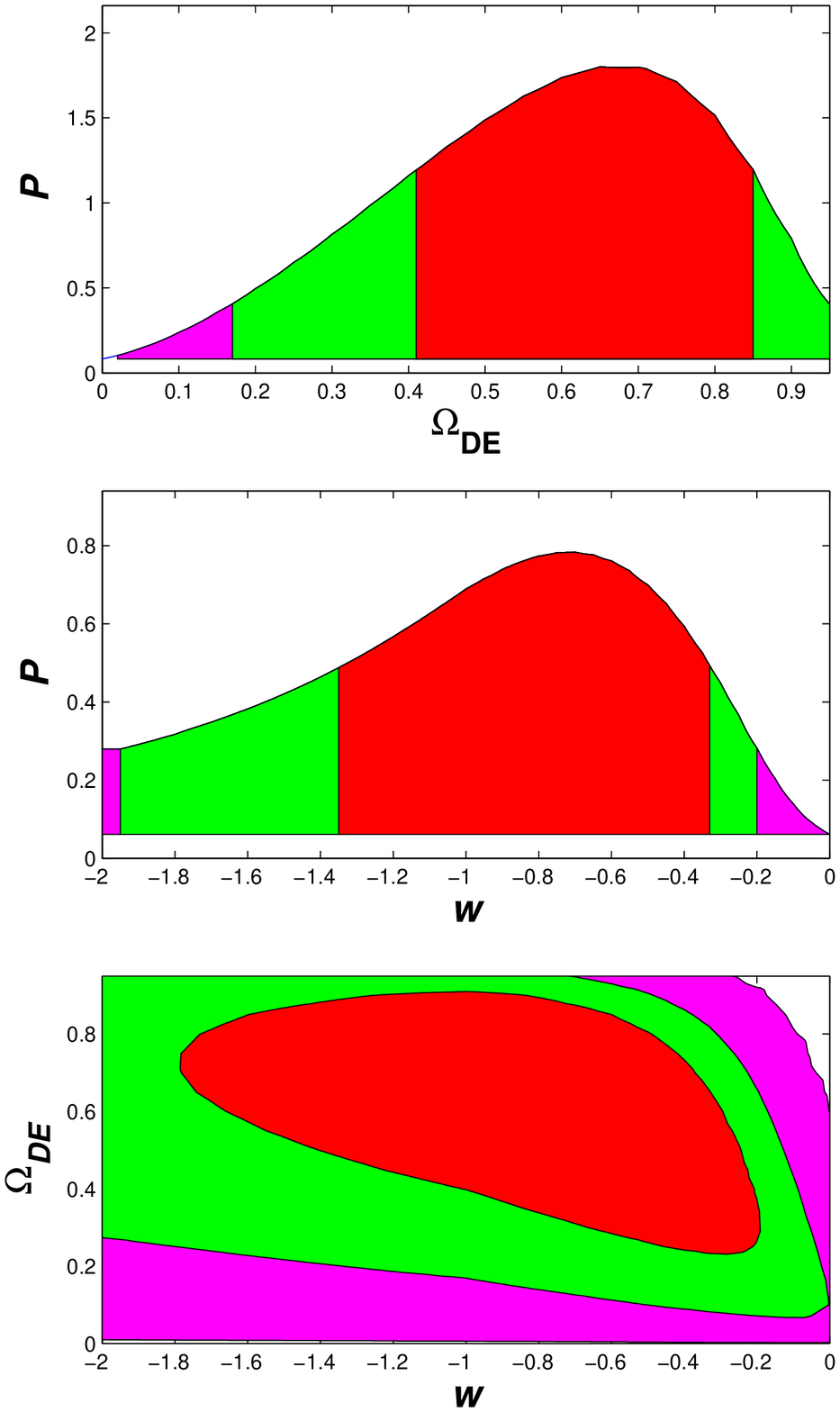}
\caption{The marginalized pdf for $\Omega_{DE}$ (upper) and $w$
(middle). Contours of the 2D likelihood are given in the lower pannel
(taken from \cite{vielva_06}). From left to right the three columns correspond to the
results obtained with the WMAP-NVSS cross-power spectrum, covariance
wavelet coefficients and temperature anisotropy--galaxy number density
cross-correlation, respectively.}  
\label{fig:ISW_like}       
\end{figure*}

Recently, an independent piece of evidence that the universe is not
Einstein-de Sitter (flat geometry with null dark energy density) has
been found by cross-correlating the CMB map with galaxy survey
maps. The same evolving gravitational potential wells which generate
the large scale structure of the galaxy distribution also produce the
gravitational redshift in the CMB photons at late times (the {\sl late ISW}
effect, see Sec.~\ref{sec:theory}). The amplitude and sign of the
cross-correlation depends on three 
parameters $\Omega_{DE}, \Omega_k$ and $h$. For a flat universe, as
indicated by many different observations as discussed above, a
positive signal will unambiguosly imply the presence of dark
energy. This is the case found recently cross-correlating the WMAP map
with different large scale structure surveys (the radiosource survey NVSS
\cite{condon_98}, the X-ray survey HEAO-1 \cite{boldt_87}, the optical SDSS
\cite{abazajian_04}, the near-infrared survey 2MASS \cite{skrutskie_97})
\cite{boughn-crittenden_04, fosalba_04, 
nolta_04, afshordi_04, vielva_06}. In \cite{vielva_06} three different
methods were used to estimate the {\sl WMAP-NVSS cross-correlation}: direct
temperature anisotropy-galaxy number density, cross-power spectrum
and covariance wavelet coefficients. A clear positive signal was found
in the three cases using a maximum likelihood analysis. The significance
of a non-null dark energy reaches $3.5\sigma$ for the
cross-power spectrum. This is also the maximum significance detection
of the ISW effect up to date (see also \cite{mcewen_06a} for a recent 
analysis using directional wavelets). The reason for the different detection
levels obtained with the three methods lies in the incomplete sky
coverage available for the analysis. Otherwise, for a whole sky
coverage the same detection level will be guaranteed by the maximum likelihood
method since the likelihood is invariant under linear transformations
of the data. In Fig.~\ref{fig:nvss_wmap} we plot the two maps together with the
combined mask needed to avoid the regions contaminated by foregrounds
in the WMAP map and unobserved regions of the NVSS survey. An analysis
allowing variations in the dark energy equation of state parameter $w$
also shows a prefered value close to $-1$. 
The 2D confidence contours for the pair $\Omega_{DE}, w$ and the
marginalized likelihoods are plotted in Fig.~\ref{fig:ISW_like}

\subsection{Summary of the main results}
\label{subsec:summary}

From the previous discussion about the cosmological implications of
the large amounts of data already collected on the CMB anisotropies,
galaxy redshift surveys, etc, the most important result is the
convergence of all those cosmological data sets towards the same model of
the universe, {\sl the concordance model}. Below we summarize the main
characteristics of this model and other consequencies implied by the
data: 

\begin{itemize}

\item The geometry of the universe is very close to flat.

\item The dynamics is dominated by a dark energy whose equation of
state is almost that of a cosmological constant.
 
\item Most of the matter content is in the form of cold dark matter.

\item The large scale structure of the universe was seeded by 
quantum fluctuations in the very early universe that evolved via
gravitational instability (the first evidence of this came from
COBE-DMR \cite{smoot_92}). 

\item The initial matter-energy density fluctuations were of the
adiabatic type. 

\item Topological defects did not play a dominant role in the
structure formation of the universe.

\item Primordial gravitational waves do not appreciably contribute to
neither the temperature nor the E-mode polarization anisotropies (up
to the present sensitivities reached).

\item The reionization happened at relatively early times,
$z\gtrsim 10$.\footnote{Considering the second release of WMAP data, $z$ is indeed around $10$.}

\item The initial fluctuations were close to a homogeneous and
isotropic Gaussian random field (see however the
next sections).   

\end{itemize}

\section{Beyond the power spectrum: sources of non-Gaussianity}
\label{sec:beyond}
\sectionmark{Beyond the power spectrum}

As it was said before, single-field inflationary models predict Gaussian 
temperature   
fluctuations homogeneously and isotropicaly distributed on the
sphere. In this case all the information is contained in the power
spectrum or, equivalently, the two-point correlation function. In the
previous sections all the discussion has been based on such
statistical quantity. However, this picture can be modified if the
origin of the primordial fluctuations is different (e.g., the presence
of topological deffects or non-standard inflation). If this is the
case, the generated fluctuations will show non-Gaussian features
(e.g., non-random phases) and the statistical information is not only
contained in the second order moment.  Even if the primordial fluctuations
generated in the early universe are Gaussian, non-Gaussianities can
appear after recombination when the non-linear matter density
evolution takes place and new anisotropies are
generated. These secondary anisotropies, as already discussed in 
Sec.~{sec:theory}, can be produced by the
gravitational lensing effect, the interaction of CMB photons with
reionised matter or the gravitational effect of the non-linear
evolution of the large scale structure.  On the other hand, the
universe is believed to be a Friedmann-Robertson-Walker (FRW) model
with a simply-connected topology. Departures from this scenario,
either its geometry or/and topology, will induce inhomogeneity or/and
anisotropy in the statistical nature of the temperature distribution
on the sphere (e.g. hot spots, rings, multiple images).  Systematics
arising from instruments and data processing as well as residues left
after foreground removal to get the intrinsic CMB signal will
inevitably introduce artefacts which will contaminate the primordial
signal. It is therefore crucial to disentangle these artifacts from
primary and secondary non-Gaussianity (see \cite{barreiro_06b} for a
recent study of the residues left by different component separation
methods).

\subsection{Models of the early universe}
\label{subsec:non-standard}
Even in the standard inflationary scenario small deviations from
normality can appear as a consequence of second-order effects during
the evolution of density fluctuations after inflation finishes (see
\cite{bartolo_04} for a detailed discussion). The non-Gaussianity
produced by those second-order effects must be therefore present in
any inflationary model. In the case of non-standard inflation, as the
curvaton scenario or inhomogeneous reheating, the primodial
non-Gaussianity should be added to that term. The level of
non-Gaussianity in the cosmological perturbations can be
phenomenologically characterized by the {\sl non-linear parameter}
$f_{NL}$ in the gravitational potential:
\begin{equation}
\Phi=\Phi_L+f_{NL}{\Phi_L}^2  ,
\end{equation}      
where $\Phi_L$ is the gravitational potential at linear
order. Although $f_{NL}$ will in general depend on the scale, from a
practical data analysis point of view it is useful to consider it as a
constant \cite{bartolo_04}. Recently, a complete second-order calculation
has allowed an estimate of an effective $f^{eff} _{NL}$ for some inflationary
scenarios and considering only a limitted number of multipoles $\ell_{max}\leq
500$ due to computational cost \cite{liguori_06}. The result is a value
$f^{eff}_{NL}(\ell_{max})\simeq 4$ with an increasing trend towards
higher multipoles. This level of non-Gaussianity might be marginally
detected with future  
experiments like Planck.

Topological defects in the universe are expected to appear as a
natural consequence of phase transitions in the early universe (for a
detailed description see \cite{vilenkin_shellard_94}). They
are associated to symmetry breaking of fields responsable for the
different particle interactions. According to their dimensionality
they can be classified in the following types: monopoles, cosmic
strings, domain walls and textures. We already know from the mismatch
between the observed oscillations of the
$C_\ell$  and the generic featureless spectra predicted by cosmic
defects that
their role in the generation of the cosmological 
fluctuations has to be sub-dominant. The contribution of both global
and local defects to the $C_\ell$ is constrained to be $\lesssim
10-20\%$ \cite{bevis_04, fraisse_05,
landriau_shellard_04, wyman_05, wu_05}. However, their intrinsically
active role in seeding  structure formation in the universe produces a
characteristic non-Gaussian CMB temperature field. On large angular
scales the superposition of many defects tends to Gaussianity by the
central limit theorem. On scales smaller than the projected
inter-defect separation non-Gaussianity can be seen as line-like
discontinuities, in the case of strings, and as hot/cold spots in the case of
monopoles and textures. Cosmic strings, the best studied defects, have
recently received renewed attention because they are produced in a
wide class of string theory models where inflation arises from the
collisions of brains.



\subsection {Geometry and topology}
\label{subsec:geometry}

The angular distribution of the CMB temperature and polarization
fluctuations on the sky is very sensitive to the spacetime metric of
the universe. On the largest scales it is assumed that the metric
corresponds to the Friedmann-Robertson-Walker one. Deviations from
this metric, even if small, can lead to noticeable signatures in the
CMB anisotropy. In particular for homogeneous and anisotropic models
(classified as Bianchi models), it is well known that global shear or
rotation can produce a spiral pattern or hot spots on the CMB sky
\cite{barrow_85, jaffe_05}. Since those signatures are on large angular scales,
strong limits on those models were already derived from the COBE-DMR
experiment \cite{kogut_97, martinez-gonzalez_sanz_95, bunn_96}. More
recently, asymmetries/non-Gaussianities found in the WMAP first-year
data have been nicely fitted with a Bianchi VII$_{\it h}$ model
\cite{jaffe_05}. However, even considering extensions of the Bianchi
VII$_{\it h}$ models in which a $\Omega_\Lambda$ term is
included, the values of the parameters which fit the large scale
asymmetries are ruled out by current observations at high significance
\cite{jaffe_06, bridges_06}.

Due to the local character of the General Theory of Relativity the
global topology of the universe is not theoretically constrained
(for different topologies discussed in the literature see e.g. 
\cite{lachieze-rey_luminet_95, weeks_98, uzan_99, levin_02}).
Non-trivial topologies leave imprints in the CMB anisotropy at large
angular scales which can be observed as anisotropic patterns, matched
circles or, more generically, deviations from a Gaussian random field
(see e.g. \cite{cornish_98, inoue_01, rocha_04}). The observability of those
signatures depends on the topology scale as compared to the horizon
scale
\cite{scannapieco_99}. Detailed CMB simulations in non-trivial
topologies are required to extract the maximum information from the
data (see \cite{riazuelo_04, hipolito-ricaldi_gomero_05}). Precision
all-sky surveys are necessary 
to obtain strong limits on global topology. The COBE-DMR data already
constrained several topological models (see
e.g. \cite{oliveira-costa_96, roukema_00, rocha_04}). More recently,
the WMAP data were also analysed to further constrain topology finding
no evidence of non-trivial topologies \cite{kunz_06,
cresswell_05}. Moreover, theoretical limits are derived for the size
of the topologies that can be detected
\cite{kunz_06}. Besides, \cite{luminet_03, roukema_04,
aurich_05} have claimed 
a dodecahedral topology for the universe to explain the anomalies
found at the lowest multipoles of the WMAP data. The next all-sky
experiment, the Planck mission, will obtain more sensitive data in a
wider frequency range allowing a better separation of the CMB from
Galactic emissions and therefore a better determination of the
low-order multipoles. It will thus help to clarify the present
situation and to determine whether the present anomalies are due to
systematics, foreground residues or are indeed true anisotropies of
the CMB.

\subsection {Secondary Anisotropies}
\label{subsec:secondary_anisotropies}

As already commented in Sec.~\ref{sec:theory}, the temperature
anisotropies of the CMB are usually divided in primary 
and secondary. The former are generated prior to recombination and are
directly related to the initial density fluctuations.  The latter are
generated after matter-radiation decoupling and arise from the
interaction of the CMB photons with the matter. These interactions can
be of gravitational type (e.g., {\sl Rees-Sciama} effect, \cite{RS,
martinez-gonzalez_90}), or of scattering type when the matter is
ionised (e.g., {\sl Sunyaev-Zel'dovich (SZ)} effect of galaxy clusters
\cite{SZ} or {\sl Ostriker-Vishniac} effect for a homogeneous re-ionisation 
\cite{OV, vishniac_87} or, on the contrary, {\sl inhomogeneous re-ionisation} of the
universe \cite{aghanim_96, gruzinov_hu_98, knox_98}).  Secondary
fluctuations will in general induce non-Gaussian signatures that will
add to the intrinsic statistical properties of the
CMB anisotropies. Their contributions to the bispectrum (third order
moment in harmonic space, see next section) during
reionization have been studied in \cite{cooray_hu_00}.  More recently,
it has been shown that the imprint of the Ostriker-Vishniac effect can
be characterized by an undetectable bispectrum and a significant
trispectrum (fourth order moment in harmonic space) \cite{castro_03}.
The Rees-Sciama effect due to 
non-linear evolution of matter is expected to produce three-point
correlations much below the cosmic variance \cite{mollerach_95}.

The SZ effect due to galaxy clusters is one of the most important
sources of secondary anisotropies. Two types of SZ effect can be
distinguished: the {\sl thermal} effect, induced by the CMB photon scattering
off free electrons in the hot intra-cluster gas, and the {\sl kinetic SZ}
effect due to Doppler shift of the photons when the clusters move with
respect to the CMB rest frame.  Owing to its peculiar spectral
signature the thermal effect will be separated to a good accuracy and
thus removed from the cosmological signal. However, the contribution
from kinetic SZ effect, which is spectrally indistinguishable from the
primary anisotropies, will not be easily subtracted, and will thus
remain a potentially significant non-Gaussian foreground
contribution. The SZ effect of galaxy clusters is by nature a
non-Gaussian process. Several studies have aimed at characterising the
non-Gaussian signature, either in wavelet space (see
e.g. \cite{aghanim_forni_99}) or in the real one (\cite{cooray_01,
yoshida_01}), in order to address its detectability and extract the
maximum information from it.

On its way from the last-scattering surface to us, the CMB radiation
passes mass inhomogeneities which deflect its paths by the
gravitational lensing effect. This process can approximately be
described as a random walk of light through a continuous field of mass
inhomogeneities, and thus leads to a diffusion process. The diffusion
tends to broaden structures in the CMB on angular scales smaller than
$\sim 10'$ (see \cite{bartelmann_schneider_01} for a review).

As long as the density inhomogeneities can be described as a Gaussian
random field, the light is slightly deflected and its distribution is
a Gaussian random process.  A purely Gaussian CMB would therefore
remain approximately Gaussian despite the deflection at least for
$l\lsim 1000$. However, at smaller scales non-Gaussianity is imposed
on an originally Gaussian CMB by producing correlations between temperature
fluctuations at large scales (through the {\sl late ISW}
effect) and light 
deflections which generate noticeble effects at several arcmin 
resolution \cite{hu_00, goldberg_spergel_99,
seljak_zaldarriaga_99}. Those correlations induce high-order
correlations in the CMB temperature and polarization fields. In
particular, a bispectrum is generated whose signal to cosmic variance
noise is higher for polarization and polarization-temperature
bispectra than for those involving the temperature alone {\cite{hu_00}}.  


Gravitational lensing by galaxy clusters can also lead to non-Gaussian
features in the CMB through their strong lensing effect. While their
origin is not physically distinct from lensing by large-scale
structures, galaxy clusters can be individually identified through
their thermal Sunyaev-Zel'dovich effect and, for the brightest cases,
extracted from CMB maps. Methods to remove non-Gaussianity from CMB
maps therefore have to distinguish between lensing by clusters and by
large-scale structures.

\section{Methods to detect and characterize deviations from
Gaussianity, standard geometry and trivial topology} 
\label{sec:methods}
\sectionmark{Methods to detect non-Gaussianity}

\begin{figure}
\centering
\includegraphics[height=6.in]{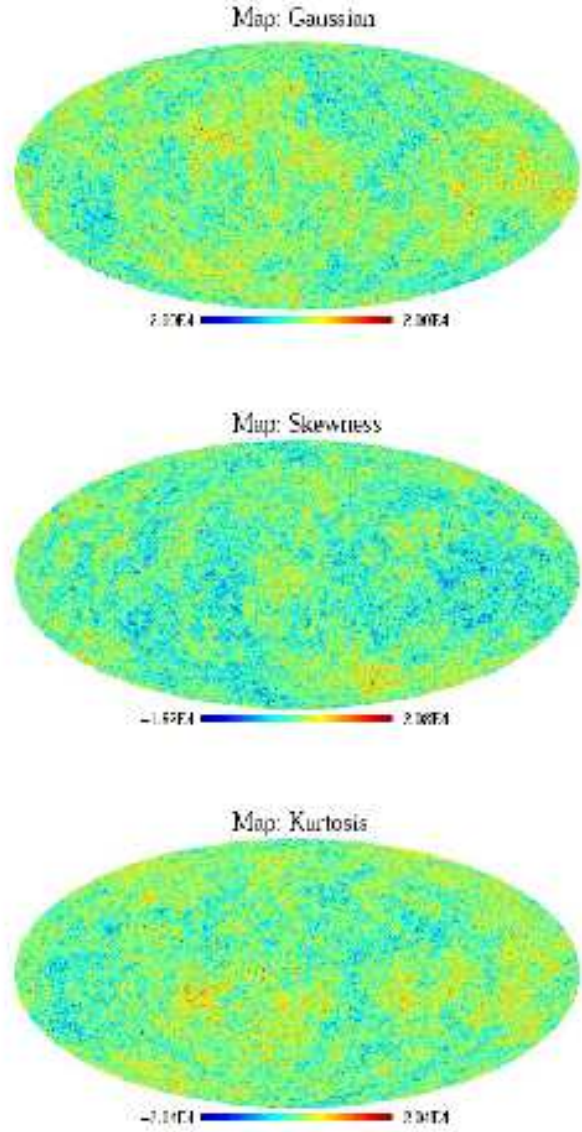}
%
%
\caption{Map simulations for Planck LFI 33\,GHz
($N_{\rm side}=256$, FWHM = 33 arcmin).  Non-Gaussian features have been
introduced in the map using the Edgeworth expansion (from
\cite{martinez-gonzalez_02}).} 
\label{fig:edgeworth}       
\end{figure}

There are infinite ways in which a random field can departure from 
Gaussianity. For instance, the cumulants above order 2 should be zero
for a Gaussian distribution and thus a non-null value taken by any of
them would represent a deviation from normality. In
Fig.~\ref{fig:edgeworth} it is shown a Gaussian map corresponding to
the concordance model together with other non-Gaussian ones with the
same power spectrum but with small amounts of skewness
or kurtosis. Hence, there is not a unique way to detect and
characterize deviations from Gaussianity. Depending on the kind of
features that are investigated some specific methods will prove to
be more efficient than others.  A variety of methods have been already
proposed to search for non-Gaussianity in the temperature or polarization
maps. Acting in different spaces (real, harmonic, wavelet, eigenmode, ...) they
are able to  
extract relevant information which is otherwise hidden in the
fluctuation maps. Improved methods for the combined
temperature and polarization maps are expected to be developed in the
coming years motivated by the large amounts of new data expected to be
collected by the new experiments (specially the all-sky WMAP and
Planck satellites).

An important difference among the methods is wether they are guided or
blind. The former are intended to test specific models of
non-Gaussianity. The latter do not assume any form of
non-Gaussianity. Guided methods would be more powerful in constraining
non-Gaussianity than the blind ones at the cost of being
model-dependent. Both methods have been used in the literature with
different results: while guided methods have imposed upper limits on
the non-linear parameter $f_{NL}$ some blind methods have obtained
significant deviations of Gaussianity (see Sec.~\ref{subsec:WMAP_NG}).   

Below we summarize different methods which can be
used to detect and characterize non-Gaussianity in
temperature/polarization maps.

Methods based on {\sl real space} are very useful for picking out
statistical features. However, they may be insensitive to any
localized features.  Standard statistics like {\sl moments} and {\sl
cumulants}, obtained from the moment generating function and its
logarithm or cumulant generating function, are included   
in the 1-point probability distribution function (pdf). Any non-zero
amplitude for any cumulant 
of order higher than 2 would be an indication of the presence of 
non-Gaussianity. It is more common to use the skewness and kurtosis, the
third and fourth standardized moments, which are defined by
the ratio of the 3 and 4-order cumulants nomalized to the square root
of the 2-order cumulant to the power of the corresponding order, respectively.
All the statistical information is, however, included in the n-point
pdf. For the Gaussian case, it is sufficient to 
specify the 2-point correlations.  Any positive
detection of reduced correlations above order 2 would imply deviations
from Gaussianity in the data. Tests of normality based on the {\sl
Edgeworth expansion} for estimating the skewness of the COBE maps have
been developed by \cite{contaldi_00}. The Edgeworth expansion allows
to describe perturbations of the Gaussian distribution in terms of a
series of cumulants (see \cite{kendall_stuart_77}). 
An alternative expansion derived from the Hilbert space of a linear
harmonic oscillator, 
producing a proper normalized distribution under 
any truncation, has been proposed by \cite{rocha_01}.
Other approaches based on {\sl goodness-of-fit tests} of normality
have recently  
been developed and applied to the MAXIMA data \cite{cayon_03a}.

A different approach to test Gaussianity is based on {\sl scalar
quantities} on the sphere. These quantities are constructed with the
first and second covariant derivatives of the temperature field. A
well known example is the local curvature (see
\cite{dore_03,hansen_04b}) but many others can be defined as the
modulus of the gradient, shape index, determinant of the Hessian
matrix, ... (see \cite{monteserin_05} for a detailed study of those
quantities). These scalar quantities have been recently
shown to be good detectors of small deviations from Gaussianity as given
by the Edgeworth expansion \cite{monteserin_06}.  Analyses based on
{\sl statistics of extrema} and {\sl excursion sets} (e.g., number,  
correlations, eccentricities) provide very robust results since extrema are 
often many sigmas above the noise level. Those quantities can be computed
(semi-)analitycally for a Gaussian field as was shown in the
pioneering work \cite{bond_efstathiou_87}.    
Also, it has been shown that the {\sl Gaussian curvature} of peaks is an 
efficient discriminator of non-Gaussianity \cite{barreiro_01}. The 
{\sl extrema correlation function} for a Gaussian field is accurately
computed in  
\cite{heavens_sheth_99, heavens_gupta_00} whereas \cite{barreiro_98}
computed the {\sl correlation of excursion sets} for Gaussian and some   
non-Gaussian temperature distributions.  {\sl Minkowski functionals}
describe the morphology of the excursion sets which are defined by
thresholding the temperature field. They are three 
for the 2D sphere: total contour of the excursion set, total area and the
total curvature (or the genus). Analytical expresions can be obtained for the 
Gaussian field. 
These morphological descriptors have
been implemented on the HEALPix pixelisation \cite{gorski_05} of the sphere by 
\cite{wandelt_98}. 
 
Alternative methods based on {\sl multifractal analysis}
and {\sl surface roughness}, widely used in many branches of physics, have
also been proposed to detect non-Gaussianity in CMB maps \cite{diego_99, 
mollerach_99}.

{\sl Spherical harmonics space} statistics are attractive as the
predictions for Gaussian fields can in some cases be computed
analytically, and their covariance properties may also be computed in
principle. In practice, we may not be able to benefit from these
advantages due to the complexities and size of the experimental data
sets.  As for the other statistics, sensitivity to artefacts such as
those arising from the observing strategy and, in particular,
incomplete foreground subtraction, need to be assessed.  The {\sl
bispectrum}, $B_{\ell_1\ell_2\ell_3}$, is the rotationally invariant
third-order moment of spherical harmonic coefficients and has been
very often used to test different data sets. It is given by the
following expression (see 
\cite{hu_01, bartolo_04} for a derivation of this and higher order
moments):
\begin{equation}
B_{\ell_1\ell_2\ell_3}= \sum_{m_1m_2m_3}\left(\begin{array}{ccc} 
\ell_1 & \ell_2 & \ell_3 \\ m_1 & m_2 & m_3 \end{array} \right) 
a_{\ell_1m_1}a_{\ell_2m_2}a_{\ell_3m_3} , 
\end{equation}
where (...) is the Wigner-3j symbol. The bispectrum must satisfy the
selection rules, requiring that $\ell_1+\ell_2+\ell_3$ be even, the
triangular rule $|\ell_1-\ell_2|\le \ell_3\le \ell_1+\ell_2$ and that
$m_3=m_1+m_2$. For 
simplicity reasons often the case that all multipoles are equal,
$\ell_1=\ell_2=\ell_3$, has been considered (see
e.g. \cite{ferreira_98} for an application to COBE-DMR). Inter-scale
correlations of the form $B_{\ell-1\ell\ell+1}$ have also been used
(e.g. \cite{magueijo_00}). They are generically expected in models with
scaling seeds like cosmic defects. Other components with more distant
multipoles may be considered but they are often dominated by noise. A
fast algorithm to compute the bispectrum from HEALPix-formatted data
has been developed by
\cite{komatsu_02} and applied to the HEALPix-formatted COBE-DMR data.   
\cite{kunz_01} have recently implemented an algorithm to compute
the {\sl trispectrum} (fourth order moment in harmonic space) in the HEALPix 
pixelization and have applied it to the COBE-DMR data.

\begin{figure}
\centering
\includegraphics[height=5.in]{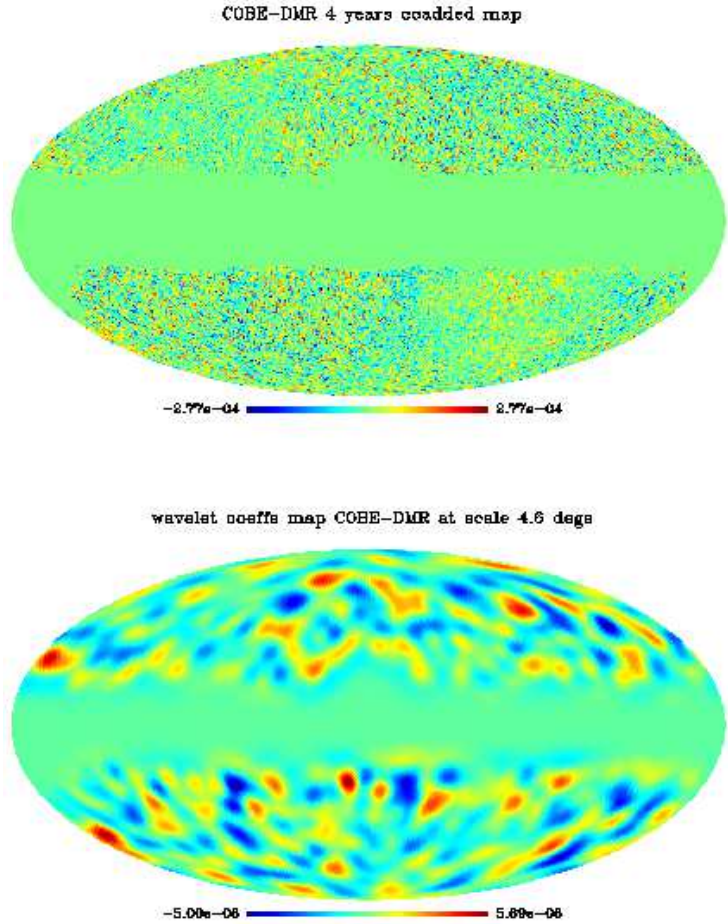}
%
%
\caption{Spherical Mexican Hat Wavelet
coeficients for COBE-DMR (from \cite{cayon_01}.}
\label{fig:SMHW_COBE}       
\end{figure}

Novel methods have been recently developed based on other spaces
where the data are transformed by either {\sl filters, wavelets or
eigenvectors}. Since the operations involved are linear the new
coefficients should form a set of Gaussian variables if the underlying
map is Gaussian.  {\sl Wavelets} are compensated filters which allow one 
to extract information which is
localized in both real and harmonic space and for this reason they can
be more sensitive than classical methods (for a detailed review of the 
properties of wavelets see \cite{jones_06}). In recent years they have
been applied to the non-Gaussian analysis of the CMB. Many of the
statistics and techniques used on real or harmonic space can be translated and
applied to wavelet space (e.g., moments, scale-scale correlations,
1-point pdf). On small patches of the sky several
{\sl planar wavelet families} have been successfully applied to detect
simulated cosmic
strings embedded in a Gaussian signal \cite{barreiro_hobson_01} and
search for discontinuities due to secondary anisotropies 
\cite{aghanim_forni_99}. 
The {\sl Haar wavelet} (the simplest wavelet that one can construct) 
\cite{tenorio_99, barreiro_00b} and the {\sl
Mexican Hat wavelet} (given by the Laplacian of a Gaussian)  
\cite{martinez-gonzalez_02, cayon_01} have been now implemented on the
sphere (see Fig.~\ref{fig:SMHW_COBE} for an application of the latter
wavelet to COBE). \cite{martinez-gonzalez_02} has shown that the
spherical Mexican Hat wavelet can be more sensitive than the spherical
Haar wavelet for detecting types of non-Gaussianity expected in some
non-standard scenarios. Moreover, the former wavelet has been found
to be more sensitive than the bispectrum for constraining the
nonlinear coupling parameter appearing in non-standard inflation using the 
COBE/DMR data 
\cite{cayon_03b}. Recently, {\sl directional wavelets} having a prefered 
direction have been
implemented on the sphere -e.g. {\sl elliptical Mexican Hat} and {\sl
Morlet}- and applied  
to the CMB analyses \cite{mcewen_05}. The
computational cost of exploring all possible directions can be greatly
reduced by using {\sl steerable wavelets} where the coefficients at any
direction can be obtained as linear combinations of the ones of a
basis with a limitted number of elements \cite{wiaux_05, wiaux_06}. 

{\sl Filters} can also help to extract non-Gaussian features from a
given data set.  
This can be done in different ways, e.g. by removing the non-cosmological 
information ({\sl Wiener}), equilizing the CMB signal ({\sl signal-whitening})
or the noise ({\sl noise-whitening}) \cite{wu_01a} or taking {\sl field
derivatives} to study discontinuities 
(for an application to cosmic strings 
see \cite{gott_90}). In the noise-whitening case the corresponding 
signal-to-noise eigenvalues allow also a compression of the data (see 
\cite{wu_01b}). 
 
In addition to the CMB temperature, the determination of the
statistical nature of polarization is crucial to test the Gaussianity
of the CMB.  Most of the tests proposed above for the temperature can
be implemented for the CMB polarization. There are, however, other
properties which are specific of the pseudo-vector nature of
polarization. Up to now only a few tests have been developed based on
{\sl geometric
characterisitics of polarization}: peculiar points, up-crossing and 
down-crossing points of the modulus of polarization (see for instance \cite{naselsky_novikov_98}). 
In any case, it would be
very desirable to device methods combining temperature and
polarization maps in an optimal way.
      
Some of the methods already discussed can as well be used to search for
possible features present in the CMB maps arising from {\sl global
anisotropy, inhomogeneity} or {\sl topology}. Particularly interesting
are those 
based on wavelets, optimal filters and pattern recognition.  If the
universe is {\sl {anisotropic/inhomoge-neous}} (as in the case of,
e.g., Bianchi, 
Tolman or Swiss-Cheese models) then the CMB temperature fluctuation field 
will be anisotropic/inhomogeneous on the sphere. The space-time
geometry of the universe is expected to affect mainly the low-order
multipoles (i.e. large angular
scales). However, focusing of geodesics can also produce a large variety of
features at smaller scales (e.g., hot spots, rings of fire, spiral
patterns, North-South asymmetries, etc., see
\cite{barrow_85,jaffe_06}. For recent 
applications to detect those features in WMAP data see
Sec.~\ref{subsec:WMAP_NG}).   
If the {\sl topology} 
of the universe is not trivial then anisotropic features will
naturally appear in the CMB maps. Also symmetries are expected in
some topologies (e.g. toroidal one). A method to probe non-trivial
topology is based on the S-statistic which measures the symmetry of a
map \cite{starobinsky_93, oliveira-costa_04}. 
Another technique consists on the {\sl identified circles} principle. 
Circular patterns of temperature fluctuations are generated in compact
topologies. A given circle shoud be seen in different 
positions of the CMB sky since it can travel to the observer
following different geodesics. Notice, however, that it will not look
exactly the same because of the ISW effect caused by different intervining 
structures and the Doppler effect at recombination whose projection on
different geodesics will be also different. This technique has been
applied to the  COBE-DMR data \cite{cornish_96,
cornish_98}. Techniques based on wavelets, that 
allow to detect the patterns of spots imprinted by the different
topologies, have been implemented in \cite{rocha_04}.  The
results show the potential of scale-scale correlations for
distinguishing among different topologies. Recently, other techniques 
involving {\sl phase correlations} have been applied to the WMAP data
to constrain the topology of the universe \cite{dineen_05}.    







\section{Data analysis beyond the power spectrum}
\label{sec:data_analysis}
\sectionmark{Data analyses beyond the power spectrum}

\subsection {The Search for non-Gaussianity before WMAP}
\label{subsec:before_WMAP}

The first systematic analyses that have been done on CMB data in search
for non-Gaussianity, have been performed on the 4 year COBE-DMR data
set. This was for a decade the only whole-sky data set publicly available. 
The COBE-DMR data have been found to be compatible with the Gaussian
hypothesis by the majority of the statistical tests applied. The few
analyses finding deviations from normality have been later proved to
be either incomplete or interpreted as undocumented systematic effects
(see below).
Monte Carlo
simulations, taking into account all the instrumental and
observational constraints of the data under analysis, 
are usually 
performed to estimate distributions of the testing-statistic as well
as confidence levels. 

The degree of asymmetry of the data was measured with the 
{\it 3-point correlation function} (equilateral and pseudo-collapsed) by 
\cite{kogut_96}. {\it Extrema distribution and correlations} of the 
temperature field have been 
tested by \cite{kogut_96, phillips_kogut_01}.   
Several works have relied on morphological characteristics of the
observed field to detect non-Gaussianity, {\it the Minkowski functionals}, 
\cite{schmalzing_gorski_98, novikov_00} and in particular the {\it
genus} (number of peaks above the threshold minus the number of
holes. \cite{kogut_96, phillips_kogut_01}) to study the COBE-DMR data.
The {\it Partition Function}, based on the combination of information
at several scales and the contribution of different moments of the
measure (defined in this case as the sum of absolute temperatures in a
defined box), and the 
roughness of the last-scattering surface were proposed by
\cite{diego_99} and \cite{mollerach_99}, respectively, as powerful
statistics to detect deviations from Gaussianity. All the analyses
performed in those works were done in real space. No evidence of
departure from Gaussianity of the COBE-DMR anisotropies was found in
all the cases discussed above.

Tests were also carried out in spaces defined by eigenmodes and
wavelets (see previous section). 
A {\it Principal Component Analysis} was done by
\cite{bromley_tegmark_99} to extract the eigenmodes from the COBE-DMR
data.  $\chi^2$, Kolmogorov-Smirnoff, cumulants up to fourth order and
significance of the top few outliers were calculated from the
eigenmodes and tested against Gaussianity. None of these tests
rejected the Gaussian hypothesis at the $95\%$ confidence
level. Skewness, kurtosis and scale-scale correlations of the COBE-DMR
{\it Wavelet} coefficients have been studied in the north and south
faces of the quad-cube COBE pixelization, using different wavelet
bases \cite{pando_98, mukherjee_00, aghanim_01}. Spherical wavelets
have also been convolved with the COBE data (HEALPix pixelization) to
test the distributions of wavelet coefficients at several scales
against Monte Carlo simulations assuming Gaussianity
\cite{barreiro_00b, cayon_01, cayon_03b}.  A detection of non-Gaussianity was
claimed by \cite{pando_98} finding the value of the scale-scale
correlation (between scales $11^{\circ}$ and $22^{\circ}$) for the
COBE data outside the $99\%$ confidence level. However,
\cite{mukherjee_00} pointed out the orientation dependence of orthogonal 
wavelet basis. 
A rotation of the data by $180^{\circ}$ makes the COBE data set
compatible with Gaussianity.

Several works have studied the {\it bispectrum} of the COBE-DMR data 
in a simplified form that ignores correlations between different multipoles
\cite{ferreira_98, heavens_98, banday_00}. A rejection of 
Gaussianity was 
claimed by \cite{ferreira_98} at the $98\%$ confidence level and appeared 
at $l=16$. In a subsequent paper
\cite{banday_00} analysed the bispectrum as a function of frequency, 
channel and time interval. The non-Gaussian signal was found to come only
from the 53 GHz map, implying a non-cosmological origin.
Removal of data corresponding to a period of time dominated by a 
systematic effect (data collected during an eclipse)  resulted in
a COBE-DMR bispectrum compatible with Gaussianity.\\
An {\it extended bispectrum} including components with correlations among 
different multipoles was studied by \cite{magueijo_00}. Gaussianity 
was rejected by the COBE-DMR data at a confidence level larger than
$98\%$. An exhaustive study of the possible non-cosmological 
origin of this detection is included in that work giving negative results.   
With the aid of the WMAP data, this detection was later proved to be due to 
pixelization artifacts of the Quad-Cube teselation of the sky and the
eclipse data.\\
All the previous works calculated the COBE-DMR bispectrum for a limited 
number of modes. More recently, \cite{komatsu_02} have measured all the 
modes available from those data, 466 in total. They conclude that the data
are consistent with Gaussianity. A recent study using the {\sl
trispectrum} also  
shows consistency with Gaussianity (having previously removed the data 
affected by the systematic effect, \cite{kunz_01}. 

While all the works mentioned above performed blind analyses, some
works tried to detect deviations from Gaussianity in the form given by
the {\sl non-linear parameter $f_{NL}$} defined in
Sec.~\ref{subsec:non-standard}. Only upper limits were
obtained at least two orders of magnitude above the values expected
in the most optimistic non-standard inflationary models (see
e.g. \cite{komatsu_02, cayon_03b} for analyses involving the bispectrum
and spherical wavelets, respectively).   
 
Uncertainty remained about whether the CMB anisotropies are Gaussian
distributed or not. Scales
observed by the COBE-DMR experiment are above the causal horizon at the 
last-scattering surface. It could well be that even if the anisotropies are
non-Gaussian distributed at smaller scales, they appear Gaussian
above degree scales due to the central limit theorem. Therefore small
angular scale observations are needed to answer this question. At present
many of these new small scale observations have been tested against
Gaussianity.  
The {\sl genus} of the data of the balloon-borne QMAP, ground-based
Saskatoon and QMASK (a combination of QMAP and Saskatoon) experiments
was compared to Gaussian predictions by \cite{park_01}. Only in the
case of QMASK was an asymmetry in the genus curve found. However this
deviation from the Gaussian genus curve was not statistically
significant. \cite{shandarin_02} calculated several {\sl morphological
quantities} from the QMASK data and concluded that they were consistent
with the values expected for a Gaussian field.
Several statistics have been computed to test
the Gaussianity of the balloon-borne MAXIMA-1 data \cite{wu_01b, santos_02,
cayon_03a, aliaga_03}. 
\cite{wu_01b} apply {\sl moments, cumulants, the Kolmogorov-Smirnof test, the 
$\chi^2$ test}, and {\sl Minkowski functionals} in eigen, real,
Wiener-filtered and signal-whitened spaces; \cite{santos_02} estimate
the {\sl bispectrum}; \cite{cayon_03a, aliaga_03} apply {\sl
goodness-of-fit} tests to the data.  Consistency with Gaussianity was
found in all cases.
The balloon observations of 
BOOMERanG maps were also searched for deviations from Gaussianity.
\cite{polenta_02} perform a real-space analysis, computing {\sl skewness, 
kurtosis} and {\sl Minkowski functionals} of those data. No
significant deviation from Gaussianity was found.

More recently and contemporary to the WMAP analyses, the statistical 
distribution of interferometric observations taken with the
VSA experiment have been also examined. \cite{savage_04} apply some
tests in real and harmonic spaces, the former to the maximum-entropy
reconstruction of the observed regions. \cite{smith_04} focus on the
{\sl bispectrum} of the data. 
\cite{rubino_06} apply {\sl smooth tests of
goodness-of-fit} developed by \cite{aliaga_05}. The results of the
analyses seem to be in relatively 
good agreement with the Gaussian assumption
(except for maybe one or two fields in the analysis of \cite{rubino_06}).

%

\subsection{WMAP and non-Gaussianity}
\label{subsec:WMAP_NG}

\begin{figure}
\centering
\includegraphics[height=2.5in, angle=0]{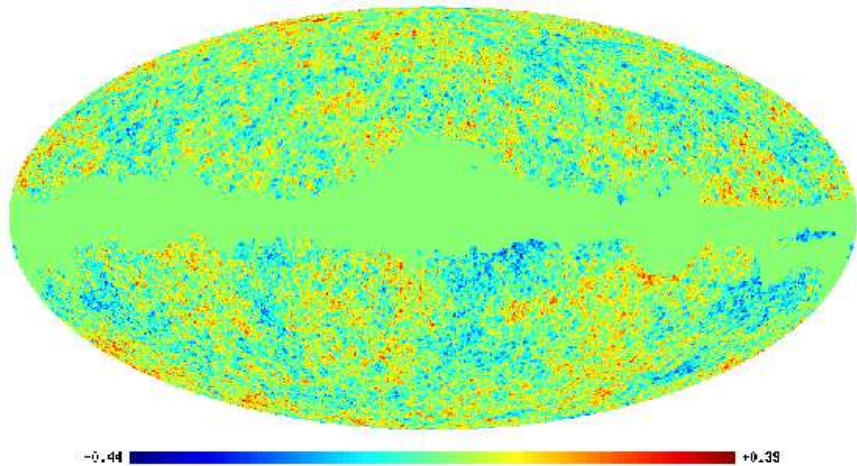}
\caption{The first year WMAP combined, foreground cleaned Q-V-W map
  together with the kp0 mask.}  
\label{fig:WCM}       
\end{figure}

The quality of the first-year all-sky data provided by the WMAP
satellite has motivated many works dealing with the Gaussianity and
isotropy of the CMB signal (also in relation to the geometry and
topology of the universe, see Sec.~\ref{subsec:geometry}). Clean maps are
constructed by the WMAP team and made publicly available to the
community, together with noise templates and beam responses for all
the channels. Also a mask is provided to ignore those pixels in the
analysis where the foreground contamination is believed to be
unavoidable. All this information is essential to perform
``realistic'' simulations of the data needed to compare the values of
the chosen statistics with the ones obtained from the true data. 
Fig.~\ref{fig:WCM} shows the first-year WMAP map obtained after a
noise-weighted combination of the foreground cleaned Q-V-W channels
(hereafter WCM). Although other foreground cleaned maps have been
produced based on different techniques \cite{bennett_03, tegmark_03,
  eriksen_04c, patanchon_05}, 
for non-Gaussianity studies the WCM (together with the most
conservative kp0 mask) is preffered since the noise
properties are well defined for this map.   

There have been a wide range of analyses of the temperature
distribution using different quantities. {\sl Minkowski functionals}
have been computed by \cite{komatsu_03, colley_gott_03, 
park_04, magueijo_medeiros_04, eriksen_04a}. The results obtained from
those works are diverse: whereas compatibility with Gaussianity is
found in \cite{komatsu_03, colley_gott_03}, others find significant
deviations from it \cite{park_04, magueijo_medeiros_04, eriksen_04a}.
The main reason for the discrepancy is that, although based on the
same quantities, they construct different measures from them as the
asymmetry in the genus above/below a given threshold or in different 
hemispheres (these latter asymmetries have also been found in the power spectrum or 
the n-point correlation 
functions for different regions of the sky, see below).\\ 
{\sl N-point correlation functions} have been calculated by
\cite{gaztanaga_wagg_03, 
eriksen_04b} for the temperature. \cite{eriksen_04b} find North-South
asymmetries on large scales in ecliptic coordinates.\\ 
Statistics based on the {\sl alignment of low-multipoles} have been computed
by \cite{oliveira-costa_04, copi_04, schwartz_04, slosar_seljak_04,
weeks_04, land_magueijo_05a,
bielewicz_05}. 
Unconfortably low probability for the observed quadrupole and octopole
alignment has been obtained in those works (the so-called axis of evil).
The {\sl bispectrum} have been calculated for different subsets of all its 
components up to the WMAP resolution \cite{komatsu_03, 
land_magueijo_05c, cabella_05b}. Significant non-null results have been 
obtained in several of those analyses which agree well with the emission
expected from undetected point sources \cite{komatsu_03, argueso_03}. 
In addition those results can be  
transtated to limits on the non-linear parameter $f_{NL}$ 
\cite{komatsu_03, cabella_05b}. By using another subset of multipole
configurations, namely the closest inter-$\ell$ components,
\cite{land_magueijo_05c} have also found North-South 
asymmetries in agreement with previous results \cite{eriksen_04b, hansen_04a}.

{\sl Phase correlations} have been used to test the hypothesis that the CMB
temperature fluctuations on the celestial sphere constitute a
homogeneous and isotropic Gaussian random field. This hypothesis
implies that the
phases of the spherical harmonic coeficients corresponding to the
temperature fluctuations are independent and uniformly distributed in
the interval [$0,2\pi$]. Phase analyses of different derived maps of
the WMAP data have been performed by
\cite{chiang_03, naselsky_03, coles_04, naselsky_05}. These works find
departures from uniformity and/or independency which in general
reflect galactic contamination or noise gradients rather than
primordial non-Gaussianity.

\begin{figure}
\centering
\includegraphics[height=4.5in, angle=270]{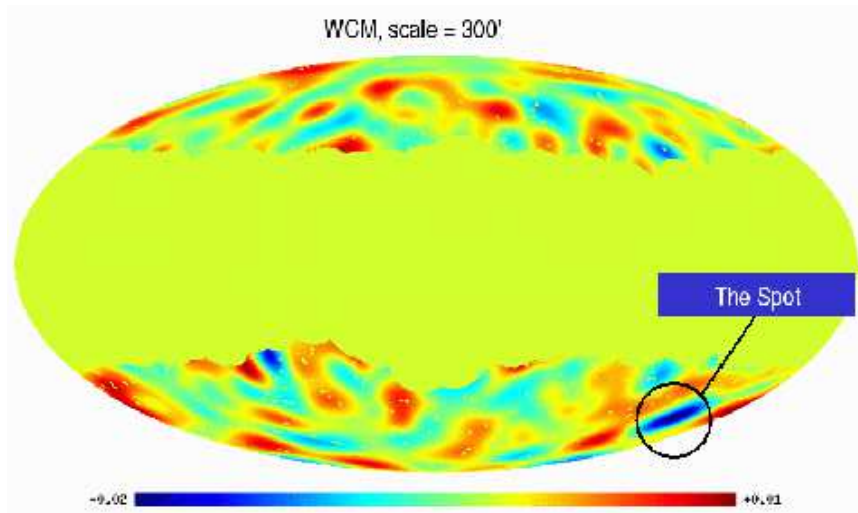}
\caption{The Cold Spot found in the WMAP combined map after convolving with a 
Spherical Mexican Hat Wavelet with scale $300'$ (see \cite{vielva_04,
cruz_05, cruz_06})}
\label{fig:spot}       
\end{figure}

\begin{figure}
\centering
\includegraphics[height=4.5in, angle=270]{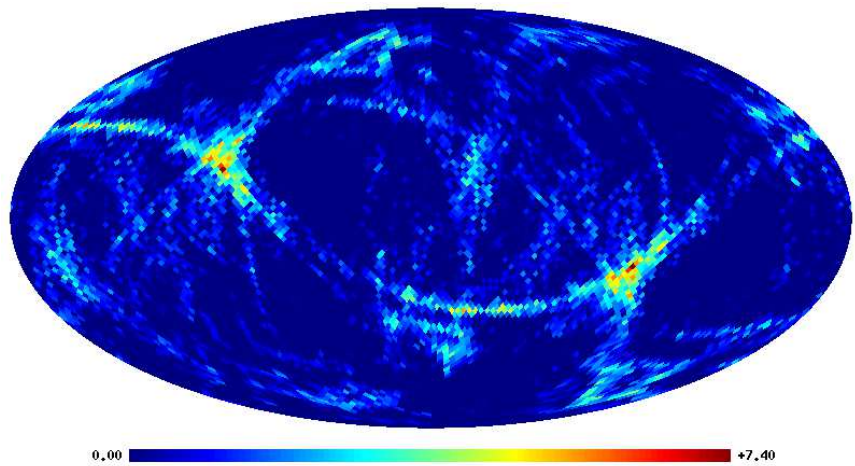}
\caption{Excess of the number of times each pixel is seen by any other
in $\sigma$ units, where $\sigma$ is the dispersion corresponding to 
Gaussian simulations (see \cite{wiaux_06})}
\label{fig:anomalous_directions}       
\end{figure}

Several statistics based on the {\sl spherical wavelet coefficients}
of the CMB data 
have been used to test the Gaussian hypothesis \cite{vielva_04, 
mukherjee_wang_04, mcewen_05, cruz_05, cayon_05,
liu_zhang_05}. Highly significant deviations have been found in the
{\sl kurtosis} of
the WMAP data for both {\sl symmetric} \cite{vielva_04, mukherjee_wang_04}
and {\sl directional wavelets} \cite{mcewen_05}. 
An analysis of {\sl extrema of
the wavelet coeficients} identifies a very {\sl cold spot} as the
cause of the large kurtosis found with the {\sl spherical Mexican Hat
wavelet} \cite{cruz_05} (see Fig.~\ref{fig:spot}). This cold spot has
been also revealed by the 
{\sl higher criticism} test applied to the wavelet coefficients
\cite{cayon_05}. Moreover, \cite{cruz_06} find a nearly spherical shape 
for the cold spot and also that it cannot be explained with 
foreground emissions.   
A test of global isotropy has been recently performed by
\cite{wiaux_06} using {\sl steerable wavelets}. Convolving the CMB
map with the steerable wavelet given by the second derivative of the
Gaussian at a fixed scale, a prefered orientation is signaled by
the maximum of the coefficients corresponding to the different
orientations and assigned to  
each pixel. Each pixel is thus looking at all the pixels intersected
by the great circle formed by extrapolating the prefered
orientation. Then the number of times each pixel is seen
by any other pixel is counted. In the case of a homogeneous
and isotropic 
Gaussian random field all the pixels are seen the same number of times
on average. Performing this test on the WMAP first-year data, a number
of anomalous directions were found at a very high significance level. 
Their mean direction is very close to the ecliptic poles, being $\approx
10^\circ$ from it (see Fig.~\ref{fig:anomalous_directions}). 
Moreover, the anomalous directions are located along a great circle
whose perpendicular is pointing towards a direction close to the
dipole and to the axis of evil. This result synthesise for
the first time the previously reported anomalies commented above on
the North-South asymmetry and the alignment of low-multipoles.\\
Contrary to these results, analyses based on the bipolar spectra do
not find violation of the global isotropy of the universe
\cite{hajian_03, hajian_05}.


{\sl Number} and {\sl correlation of extrema} are computed by
\cite{larson_wandelt_04, 
larson_wandelt_05, tojeiro_06}. \cite{larson_wandelt_04,
larson_wandelt_05} find that the maxima and minima in the WMAP data
are not hot and cold enough. \cite{tojeiro_06} find evidence for
non-Gaussianity on large scales which could be associated with
residual foregrounds.
  
{\sl Local curvature} has been investigated by \cite{hansen_04b,
cabella_05a}. They classify the points in the data as hills, lakes or
saddles according to the sign of the eigenvalues of the Hessian matrix, and
consider the proportions of each of them above a certain
threshold. A clear asymmetry is found which is maximum for hemispheres
centered near the ecliptic poles, consistent with the results by
\cite{eriksen_04b}.

Several suggestions have been made to explain the anomalies found in
the first-year WMAP data at the large angular scales, namely the power
deficit and alignment of the low-order multipoles, the asymmetries
between different hemispheres and the cold spot\footnote{These
  anomalies remained almost unaltered in the 3-years WMAP data.}.
 
A suggestion to
explain the low-multipole anomaly invokes a non-trivial topology which
due to the finite size of the universe tends to cut power at the
lowest multipoles and in some cases to align them \cite{luminet_03, 
roukema_04, aurich_05}. However, apart from this nice consequence,
there is no evidence for a possible non-trivial topology. Another
possibility proposed is the lensing of the 
dipole by local large-scale structures which can potentially produce a
quadrupole and octopole similar to those found in the WMAP data 
\cite{vale_05}. It remains to be demonstrated that the appropriate
physical characteristics of those structures (in terms of mass, size
and position) are in fact present in the local universe.  

Possible explanations for the cold spot are the RS effect produced by
the most prominent voids or massive structures in the universe
\cite{martinez-gonzalez_sanz_90, tomita_05, inoue_silk_06},
topological defects in the form or textures \cite{turok_spergel_90}
or, a more radical one, based on 
an inhomogeneous model of the universe given by a finite sphere of
dust and dark energy \cite{adler_06}.

Anisotropic and inhomogeneous geometries leaving spiral or stripy
patterns in the CMB sky have been proposed to correct the large-angle
anomalies found (asymetries, the cold spot and the low-multipole power
deficit and alignment) \cite{jaffe_05, land_magueijo_05b, cayon_06, bridges_06}
(see however 
\cite{mcewen_06b} for further non-Gaussian detections in the Bianchi
VII$_{\it h}$ corrected WMAP first year data). Although a good fit can
be found for most of the 
anomalies with the anisotropic geometries (Bianchi VII$_{\it h}$) however the
values required for the matter content are too high
\cite{jaffe_06,bridges_06}. 
 
Small changes in the magnitude of the WMAP
dipole vector \cite{freeman_06}, together with possible systematics or
foreground 
residuals associated with the ecliptic (as might be infered from the
recent results by \cite{eriksen_04b, hansen_04a, wiaux_06}), might be
considered as a possibility to explain the large-scale 
anomalies. However much more work is still needed in order to find a
convincing case for them.



\section{Conclusions}
\label{sec:conclusions}
\sectionmark{Conclusions}

Cosmology is passing through a period of abundance of new data which
imply a significant improvement on the precission in the determination
of the parameters characterizing the universe. The CMB field is
probably the major contributor to that improvement and is expected to
be so in the next decade. With only the first year of WMAP data , the
power spectrum of the temperature   
anisotropies has been determined with error bars below cosmic variance
up to multipoles $\ell \approx 600$. Information on larger multipoles
is already provided by several high-resolution experiments allowing a
consistent determination of the third acoustic oscillation and
providing an incomplete coverage of the multipoles up to $\ell \approx
2500$. In the next few years, the ESA Planck mission is expected to provide
accurate and high 
resolution all-sky images of the CMB temperature anisotropies in a
wide frequency range which will allow an optimal removal of foregrounds
and a very good control of systematics. From those high quality images
a complete coverage of multipoles at the cosmic
variance limit will be obtained up to $\ell \approx 2500$. The
$C_{\ell}$ so obtained will be the ultimate temperature power
spectrum of the primary CMB anisotropies since secondary anisotropies
are expected to start to be dominant beyond those multipoles. 

The large-scale galaxy surveys as well as the high-z SN
Ia light curves have also played a crucial role in establishing the
concordance model. The strong complementarity of these two
cosmological data sets with the CMB anisotropies has implied a
significant improvement on the parameter
determination compared to each data set considered individually. And,
what is more important, a consistency check of the best fit
model. In the next years, that combination of data sets is expected to
be strengthened with weak lensing observations of background galaxies in
large areas of the sky.    

In addition to the power spectrum, the recent studies on the statistical 
distribution of the temperature anisotropies provide independent tests
of the standard inflationary model. Significant deviations from a
homogeneous and isotropic Gaussian random field have already been
found in the first-year WMAP data. However the origin of those
deviations, whether cosmological, residual foregrounds or systematics,
is still a matter of debate. More precised maps from next years of
WMAP data and specially the high quality maps expected from the
Planck mission will certainly help in the clarification of the
possible causes of the anomalies found.

Finally, high sensitive polarization maps to search for the B-mode
will be the next major quest after Planck. The B-mode polarization is
at present the best way to study the primordial background of
gravitational waves expected in the standard model of inflation. It is
therefore not surprising that both space
agencies, NASA and ESA, have considered polarization missions in their
future programs ``Beyond Einstein'' and ``Cosmic Vision'',
respectively.

\vskip 1.truecm

{\noindent{\bf Acknowledgments}}\\
I thank J.L. Sanz and P. Vielva for useful comments on the manuscript.  I
acknowledge financial support from the Spanish MEC project
ESP2004-07067-C03-01.  I also acknowledge the use of LAMBDA, support
for which is provided by by the NASA Office of Space Science. I
acknowledge the use of the software package CMBFAST
(http://www.cmbfast.org) developed by Seljak and Zaldarriaga. The work
has also used the software package HEALPix
(http://www.eso.org/science/healpix) developed by K.M. Gorski,
E.F. Hivon, B.D. Wandelt, J. Banday, F.K. Hansen and M. Barthelmann.

\printindex

\begin{thebibliography}{99.}
%
%
%




\bibitem{abazajian_04} Abazajian K. et al., 2004, AJ, 128, 502
%
\bibitem{adler_06} Adler R.J., Bjorken J.D. \& Overduin J.M., 2006,
  gr-qc/0602102
%
\bibitem{aghanim_96} Aghanim N., Desert F.X., Puget J.L. \& 
Gispert R., 1996, A\&A, 311, 1.
%
\bibitem{aghanim_forni_99} Aghanim N., Forni O. 1999, A\&A, 347, 409
%
\bibitem{aghanim_01} Aghanim N., Forni O. \& Bouchet F.R., 2001, A\&A,
365, 341-346 
%
\bibitem{aliaga_03} Aliaga A.M., Mart\'\i nez-Gonz\'alez E., Cay\'on
  L., Arg\"ueso F., Sanz J.L. \& Barreiro R.B., 2003, New
  Astron. Rev., 47, 821
%
\bibitem{aliaga_05} Aliaga A.M., Rubi\~no-Mart\'\i n J.A., Mart\'\i
  nez-Gonz\'alez E., Barreiro R.B. \& Sanz J.L., 2005, MNRAS, 356, 1559 
%
\bibitem{afshordi_04} Afshordi N., Loh Y.-S. \& Strauss M. A., 2004, Phys. Rev. D, 69, 3524
%
\bibitem{allen_02} Allen A.W., Schmidt R.W. \& Fabian A.C., 2002, MNRAS, 334, L11
%
\bibitem{argueso_03} Arg\"ueso F., Gonz\'alez-Nuevo J. \& Toffolatti
L., 2003, ApJ, 598, 86-96
%
\bibitem{aurich_05} Aurich R., Lustig S. \& Steiner F., 2006, MNRAS,
  369, 240, astro-ph/0510847
%
\bibitem{banday_00} Banday A.J., Zaroubi S. \& Gorski K.M., 2000,
ApJ, 533, 575-587 
%
\bibitem{barreiro_98} Barreiro R.B., Sanz J.L.,
Mart\'{\i}nez-Gonz\'alez E. \& Silk J., 1998, MNRAS, 296, 693
%
%
\bibitem{barreiro_00b} Barreiro R.B., Hobson M.P., Lasenby A.N.,
Banday A.J., Gorski K.M. \& Hinshaw G., 2000, MNRAS, 318, 475-481
%
\bibitem{barreiro_hobson_01} Barreiro R.B. \& Hobson M.P., 2001,
MNRAS, 327, 813, astro-ph/0104300
%
\bibitem{barreiro_01} Barreiro R.B., Mart\'{\i}nez-Gonz\'alez E.\& Sanz J.L. 2001, MNRAS, 322, 411
%
\bibitem{barreiro_06a} Barreiro R.B., 2006, in this volume, astro-ph/0512538
%
\bibitem{barreiro_06b} Barreiro R.B., Mart\'{\i}nez-Gonz\'alez E., Vielva P. \& 
Hobson M.P., 2006, MNRAS, 368, 226
%
\bibitem{barrow_85} Barrow J.D., Juszkiewicz R. \& Sonoda D.H., 1985,
  MNRAS, 213, 917
%
%
\bibitem{bartelmann_schneider_01} Bartelmann M. \& Schneider
P., 2001, Phys. Rep. 340, 291 
%
\bibitem{bartolo_04} Bartolo N., Komatsu E., Matarrese S. \& Riotto A., 2004, Phys. Rep., 402, 103
%
\bibitem{bennett_03} Bennett C.L. et al., 2003, ApJS, 148, 1
%
\bibitem{bersanelli_02} Bersanelli M., Maino, D. \& Mennella A., 2002, Nuovo Cimento, 25, 1
%
\bibitem{bevis_04} Bevis N., Hindmarsh M. \& Kunz M., 2004,
Phys. Rev. D, 70, 043508
%
\bibitem{bielewicz_05} Bielewicz P., Eriksen H.K., Banday A.J.,
  G\`orski K.M. \& Lilje P.B., 2005, ApJ, 635, 750 
%
\bibitem{boldt_87} Boldt E., 1987, Phys. Rev., 146, 215
%
\bibitem{bond_efstathiou_87} Bond J.R. \& Efstathiou G., 1987, MNRAS,
  226, 655 
%
\bibitem{bond_05} Bond J.R. et al., 2005, ApJ, 626, 12
%
\bibitem{bouchet_99} Bouchet F.R. \& Gispert R., 1999, New Astronomy Reviews, 4, 443
%
\bibitem{bouchet_05} Bouchet F.R., Beno\^\i t A., Camus Ph., D\'esert
F.X., Piat M. \& Ponthieu N., 2005: Charting the New Frontier of the
Cosmic Microwave Background Polarization. In: \textit{SF2A}, ed. by F.
Casoli et al. (EDP Sciences 2004), astro-ph/0510423
%
\bibitem{boughn-crittenden_04} Boughn S.P. \& Crittenden R.G., 2004,
Nature, 427, 45 
%
\bibitem{bridges_06} Bridges M., Lasenby A.N. \& Hobson M.P., 2006,
  MNRAS Letters, submitted, astro-ph/0605325
%
\bibitem{bromley_tegmark_99} Bromley B.C. \& Tegmark M., 1999, ApJ,
524, L79-L82. 
%
\bibitem{bucher_01} Bucher M., Moodley K. \& Turok N., 2001,
Phys. Rev. Lett., 87, 191301
%
\bibitem{bunn_96} Bunn, E.F., Ferreira, P. \& Silk, J. 1996, Phys. Rev. Lett., 77, 2883
%
\bibitem{bunn_03} Bunn E.F., Zaldarriaga M., Tegmark M. \& de
Oliveira-Costa A., 2003, Phys. Rev. D, 67, 023501
%
\bibitem{cabella_04} Cabella, P. \& Kamionkowski, M., 2004, Theory of
  Cosmic Microwave Background Polarization. In: \textit{The
  Polarization on the Cosmic Microwave Background}, 2003 Villa
  Mondragone School of Gravitation and Cosmology, Rome, September 6-11,
  astro-ph/0403392
%
\bibitem{cabella_05a} Cabella P., Liguori M., Hansen F.K., Marinucci D., 
Matarrese S., Moscardini L. \& Vittorio N., 2005, MNRAS, 358, 684
%
\bibitem{cabella_05b} Cabella P., Hansen F.K., Liguori M., Marinucci D., 
Matarrese S., Moscardini L. \& Vittorio N., 2006, MNRAS, 369, 819,
astro-ph/0512112 
%
\bibitem{castro_03} Castro P.G., 2003, Phys. Rev. D, 67, 123001,
  astro-ph/0212500 
%
\bibitem{cayon_01} Cay\'on L., Sanz J.L., Mart\'\i nez-Gonz\'alez E., 
Banday A.J., Arg\"ueso F., Gallegos J.E., Gorski K.M. \& Hinshaw G., 2001,
MNRAS, 326, 1243
%
\bibitem{cayon_03a} Cay\'on L., Arg\"ueso F., Mart\'\i nez-Gonz\'alez
E. \& Sanz, 2003, MNRAS, 344, 917
%
\bibitem{cayon_03b} Cay\'on L., Mart\'\i nez-Gonz\'alez E., Arg\"ueso F., 
Banday A.J., Gorski K.M., 2003, MNRAS, 339, 1189
%
\bibitem{cayon_05} Cay\'on L., Jin L. \& Treaster A., 2005, MNRAS, 362, 826
%
\bibitem{cayon_06} Cay\'on L., Banday A.J., Jaffe T., Eriksen H.K.,
  Hansen F.K., Gorski K.M. \& Jin J., 2006, MNRAS, 369, 598  
%
\bibitem{challinor_04} Challinor A., 2004, Anisotropies in the Cosmic
  Microwave Background. In: \textit{Proceedings of the 2nd
  Aegean Summer School on the Early Universe}, 22-30 September 2003,
  to appear in Springer LNP, astro-ph/0403344 
%
\bibitem{challinor_05} Challinor A., 2006, in this volume, astro-ph/0502093
%
\bibitem{challinor_chon_05} Challinor A. \& Chon G., 2005, MNRAS, 360,
509
%
\bibitem{chiang_03}{Chiang L.-Y., Naselsky P.D., Verkhodanov O.V. \& Way
M.J., 2003, ApJL, 590, 65}
%
\bibitem{coles_04} Coles P., Dineen P., Earl J. \& Wright D., 2004,
  MNRAS, 350, 989 
%
\bibitem{colley_gott_03} Colley W.N. \& Gott J.R., 2003, MNRAS, 344, 686 
%
\bibitem{condon_98}{Condon J.J. et al., 1998, AJ, 115, 1693}
%
\bibitem{contaldi_00}{Contaldi C.R., Ferreira P.G. \&
Magueijo J., 2000, ApJ, 534, 25}
%
\bibitem{cooray_hu_00} Cooray, A.R. \& Hu, W. 2000, ApJ, 534, 533
%
\bibitem{cooray_01} Cooray, A. 2001, Phys. Rev., D64, 063514, astro-ph/0105063
%
\bibitem{copi_04} Copi C.J., Huterer D. \& Starkman G.D., 2004,
  Phys. Rev. D, 70, 043515, astro-ph/0310511
%
\bibitem{cornish_96} Cornish N.J., Spergel D.N. \&  Starkman G.D.,
  1996, Phys. Rev. Lett., 77, 215  
%
\bibitem{cornish_98} Cornish N.J., Spergel D.N. \& Starkman G.D.,
1998, Class. Quant. Grav., 15, 2657-2670
%
\bibitem{cresswell_05} Cresswell J.G., Liddle A.R., Mukherjee P. \&
Riazuelo A., 2006, Phys. Rev. D, 73, 041302, astro-ph/0512017
%
\bibitem{croft_02}{Croft R.A.C. et al., 2002, ApJ, 581, 20}
%
\bibitem{crotty_03}{Crotty P., Lesbourgues J. \& Pastor S., 2003,
Phys. Rev. D, 67, 123005}
%
\bibitem{cruz_05} Cruz M., Mart\'\i nez-Gonz\'alez E., Vielva P. \&
  Cay\'on L., 2005, MNRAS, 356, 29 
%
\bibitem{cruz_06} Cruz M., Tucci M., Mart\'\i nez-Gonz\'alez E. \&
  Vielva P., 2006, MNRAS, 369, 57
%
\bibitem{de bernardis_00} {de Bernardis P. et al., 2000, Nature, 404,
955}
%
\bibitem{oliveira-costa_96} de Oliveira-Costa A., Smoot G.F. \&
Starobinsky A.A., 1996, ApJ, 468, 457
%
\bibitem{oliveira-costa_04}{de Oliveira-Costa A., Tegmark M., Zaldarriaga M. \& Hamilton A., 2004, Phys. Rev. D, 69, 063516}
%
\bibitem{delabrouille_06} Delabrouille R.B., 2006, in this volume.
%
\bibitem{delabrouille_03}{Delabrouille J., Cardoso J.-F. \& Patanchon G., 2003, MNRAS, 
346, 1089}
%
\bibitem{dickinson_04}{Dickinson C. et al., 2004, MNRAS, 353, 732}
%
\bibitem{diego_99}{Diego J.M., Mart\'{\i}nez-Gonz\'alez E., Sanz J.L.,
Mollerach S. \& Mart\'{\i}nez V.J., 1999, MNRAS, 306, 427-436}
%
%
\bibitem{dineen_05} Dineen P., Rocha G. \& Coles P., 2005, MNRAS, 358, 1285
%
\bibitem{dore_03} Dor\'e O., Colombi S. \& Bouchet F.R., 2003, MNRAS,
  344, 905
%
\bibitem{efstathiou_04}{Efstathiou G., 2004, MNRAS, 348, 885}
%
%
\bibitem{eriksen_04a}{Eriksen H.K., Novikov D.I., Lilje P.B., Banday
A.J., Gorski K.M., 2004, ApJ, 612, 64} 
%
\bibitem{eriksen_04b}{Eriksen H.K., Hansen F.K., Banday A.J., Gorski K.M. \&
Lilje P.B., 2004, ApJ, 605, 14}
%
\bibitem{eriksen_04c} Eriksen H.K., Banday A.J., Gorski K.M. \& Lilje
  P.B., 2004, ApJ, 612, 633
%
\bibitem{ferreira_98}{Ferreira P.G., Magueijo J. \& Gorski K.M.,
1998, ApJ, 503, L1-L4} 
%
\bibitem{fosalba_04} {Fosalba P., Gazta\~naga E. \& Castander F., 2004, ApJL, 
597, 89} 
%
\bibitem{fraisse_05}{Fraisse A.A., 2005, Phys. Rev. Lett., submitted, astro-ph/0503402}
%
\bibitem{freeman_06} Freeman, P.E., Genovese, C.R., Miller, C:J., Nichol, R.C. \& Wasserman, L., 2006, ApJ, 638, 1  
%
\bibitem{freedman_01}{Freedman W.L. et al., 2001, ApJ, 553, 47}
%
\bibitem{gaztanaga_98} Gazta\~naga E., Fosalba P. \& Elizalde
E., 1998, MNRAS, 295, L35-L39 
%
\bibitem{gaztanaga_wagg_03} Gazta\~naga E. \& Wagg J., 2003,
Phys. Rev. D, 68, 021302
%
\bibitem{goldberg_spergel_99} Goldberg D.M. \& Spergel D.N., 1999,
Phys. Rev. D59, 103002. 
%
\bibitem{gorski_05} Gorski K.M., Hivon E., Banday A.J., Wandelt B.D.,
Hansen F.K., Reinecke M. \& Bartelman M. 2005, ApJ, 622, 759 
%
\bibitem{gott_90} Gott J. R. III, Park, C., Juszkiewicz, R., Bies,
William E., Bennett, D. P., Bouchet, F.R., Stebbins, A. 1990, ApJ, 352, 1
%
\bibitem{graham_95} Graham P., Turok N. Lubin P.M. \& Schuster
J.A., 1995, ApJ, 449, 404-412 
%
\bibitem{gruzinov_hu_98} Gruzinov, A. \& Hu, W. 1998, ApJ, 508, 435.
%
\bibitem{hajian_03} Hajian A. \& Souradeep T., 2003, ApJ, 597, 5
%
\bibitem{hajian_05} Hajian A., Souradeep T. \& Cornish N., 2005, ApJ, 618, 63
%
\bibitem{hanany_00}{Hanany S. et al., 2000, ApJ, 545, L5}
%
\bibitem{hansen_04a} Hansen F.K., Banday A.J. \& Gorski K.M., 2004, MNRAS,
354, 641
%
\bibitem{hansen_04b} Hansen F.K., Cabella P., Marinucci D. \& Vittorio
  N., 2004, ApJ, 607, L67
%
\bibitem{hawkins_03} Hawkins E. et al., 2003, MNRAS, 346, 78
%
\bibitem{heavens_98} Heavens A.F., 1998, MNRAS, 299, 805-808.
%
\bibitem{heavens_sheth_99}{Heavens A.F. \& Sheth R.K., 1999, MNRAS, 310, 1062}
%
\bibitem{heavens_gupta_00}{Heavens A.F. \& Gupta S., 2000, MNRAS, 324, 960, astro-ph/0010126}
%
\bibitem{herranz_02}{Herranz D., Sanz J.L., Hobson M.P., Barreiro R.B., Diego J.M., Mart\'\i nez-Gonz\'alez E. \& Lasenby A.N., 2002, MNRAS, 336, 1057}
%
\bibitem{hinshaw_03}{Hinshaw G. et al., 2003, ApJ, 148, 135}
%
\bibitem{hipolito-ricaldi_gomero_05} Hip\'olito-Ricaldi W.S. \& Gomero
G.I., 2005, Phys. Rev. D, 72, 103008
%
\bibitem{hobson_98}{Hobson M.P., Jones A.W., Lasenby A.N. \& Bouchet F., 1998, 
MNRAS, 300}
%
\bibitem{hu_00} Hu W., 2000, Phys. Rev. D, 62, 043007
%
\bibitem{hu_01} Hu W., 2001, Phys. Rev. D, 64, 083005
%
\bibitem{hu_dodelson_02}{Hu W. and Dodelson S., 2002, ARAA, 40, 171}
%
\bibitem{hu_okamoto_02} Hu W. \& Okamoto T., 2002, ApJ, 574, 566
%
\bibitem{inoue_01} Inoue K.T., 2001, Prog. Theor. Phys., 106, 39,
astro-ph/0102222 
%
\bibitem{inoue_silk_06} Inoue K.T. \& Silk J., 2006, ApJ, 648, 23
%
\bibitem{jaffe_05}{Jaffe T.R., Banday A.J., Eriksen H.K., Gorski K.M. \& 
Hansen F.K., 2005, ApJL, 629, L1, astro-ph/0503213}
%
\bibitem{jaffe_06}{Jaffe T.R., Hervik S., Banday A.J., Gorski K.M.,
2006, ApJ, 644, 701, astro-ph/0512433} 
%
\bibitem{jassal_05}{Jassal H.K., Bagla J.S. \& Padmanabhan T., 2005, astro-ph/0506748}
%
\bibitem{jones_06} Jones B.J.T., 2006, in this volume
%
\bibitem{jones_05} Jones W.C. et al., 2006, ApJ, 647, 823, astro-ph/0507494
%
\bibitem{kamionkowski_97}{Kamionkowski M., Kosowsky A. \& Sttebins A., 1997, 
Phys. Rev. D, 55, 7368}
%
\bibitem{kendall_stuart_77} Kendall M.G. \& Stuart A., 1977, The
  Advance Theory of Statistics, London: Charles Griffin
%
\bibitem{kesden_03} Kesden M., Cooray A. \& Kamionkowski M., 2003,
Phys. Rev. D, 67, 123507
%
\bibitem{kinney_02}{Kinney W.H., 2002, Phys. Rev. D, 66, 083508}
%
\bibitem{knop_03}{Knop R.A. et al., 2003, ApJ, 598, 102}
%
\bibitem{knox_98} Knox, L., Scoccimarro, R. \& Dodelson, S. 1998, Phys. Rev. 
Lett., 81, 2004
%
\bibitem{knox_02} Knox L. \& Song Y.-S., 2002, Phys. Rev. Lett., 89, 011303
%
\bibitem{kogut_96} Kogut A., Banday A.J., Bennett C.L., Gorski K.M.,
Smoot G.F. \& Wright E.L., 1996, ApJ, 464, L29-L33. 
%
\bibitem{kogut_97} Kogut A., Hinshaw G. \& Banday A.J., 1997,
Phys. Rev. D, 55, 1901
%
\bibitem{kogut_03}{Kogut A. et al., 2003, ApJS, 148, 161}
%
\bibitem{kovac_02}{Kovac J.M. et al. 2002, Nature, 420, 772}
%
\bibitem{komatsu_02}{Komatsu E., Wandelt B.D., Spergel D.N., Banday
A.J. \& Gorski K.M., 2002, ApJ., 566, 19}
%
\bibitem{komatsu_03} {Komatsu E. et al., 2003, ApJS, 148, 119}
%
\bibitem{kunz_01}{Kunz M., Banday A.J., Castro P.G., Ferreira
P.G. \& Gorski K.M., 2001, ApJ, 563, 99}
%
\bibitem{kunz_06} Kunz M., Aghanim N., Cay\'on L., Forni O. \&
Riazuelo A., 2006, Phys. Rev. D, 73, 023511, astro-ph/0510164
%
\bibitem{kuo_04}{Kuo C.L. et al., 2004, ApJ, 600, 32}
%
\bibitem{lachieze-rey_luminet_95} Lachi\`eze-Rey, M. \& Luminet, J.-P. 1995,
    Phys.  Rep.,  254, 135
%
\bibitem{land_magueijo_05a}{Land K. \& Magueijo J., 2005, Phys. Rev. Lett., 
95, 071301}
%
\bibitem{land_magueijo_05b} Land K. \& Magueijo J., 2005, MNRAS, 367, 1714,
astro-ph/0509752
%
\bibitem{land_magueijo_05c} Land K. \& Magueijo J., 2005, MNRAS, 357,
994
%
\bibitem{landriau_shellard_04} Landriau M. \& Shellard E.P.S., 2004,
  Phys. Rev. D, 69, 023003
%
\bibitem{larson_wandelt_04}{Larson D.L. \& Wandelt B.D., 2004, ApJ, 613, 85}
%
\bibitem{larson_wandelt_05}{Larson D.L. \& Wandelt B.D., 2005, Phys. Rev. D, submitted, astro-ph/0505046}
%
\bibitem{leitch_04}{Leitch E.M., Kovac J.M., Halverson N.W., Carlstrom
  J.E., Pryke C. \& Smith M.W.E., 2005, ApJ, 624, 10, astro-ph/0409357}
%
\bibitem{levin_02} Levin J., 2002, Phys. Rep., 365, 251-333
%
\bibitem{lewis_00} Lewis A., Challinor A. \& Lasenby A., 2000, ApJ,
538, 473, http://camb.info
%
\bibitem{lewis_02} Lewis A., Challinor A. \& Turok N., 2002,
Phys. Rev. D, 65, 023505
%
\bibitem{lewis_04} Lewis A., 2004, Phys. Rev. D, 70, 043011
%
\bibitem{liddle_lyth_00} Liddle A.R. \& Lyth D.H., 2000,
\textit{Cosmological Inflation and Large-Scale Structure}, Cambridge
University Press
%
\bibitem{liguori_06} Liguori M., Hansen F.K., Komatsu E., Matarresse
  S. \& Riotto A., 2006, Phys. Rev. D, 73, 043505
%
\bibitem{liu_zhang_05} Liu X. \& Zhang S.N., 2005, ApJ, 633, 542
%
\bibitem{luminet_03}{Luminet J.-P., Weeks J.R., Riazuelo A., Lehoucq R. \&
Uzan J.-P., 2003, Nature, 425, 593}
%
\bibitem{lyth_wands_02} Lyth D.H. \& Wands D., 2002, Phys. Lett. B, 524, 5
%
\bibitem{clover} Maffei B. et al., 2005, EAS Publications Series,
14, 251-256
%
\bibitem{magueijo_00}{Magueijo J., 2000, ApJ, 528, L57-L60}
%
\bibitem{magueijo_medeiros_04}{Magueijo J. \& Medeiros J., 2004, MNRAS, 351, L1}
%
\bibitem{maino_02}{Maino D. et al., 2002, MNRAS, 334,53}
%
\bibitem{martinez-gonzalez_90}{Mart\'\i nez-Gonz\'alez E., Sanz
  J.L. \& Silk J., 1990, ApJL, 335, L5}
%
\bibitem{martinez-gonzalez_sanz_90}{Mart\'\i nez-Gonz\'alez E. \& Sanz
  J.L., 1990, MNRAS, 247, 473}
%
\bibitem{martinez-gonzalez_sanz_95} Mart\'{\i}nez-Gonz\'alez, E. \& Sanz, J.L. 1995, A\&A, 300, 346
%
\bibitem{martinez-gonzalez_02}{Mart\'{\i}nez-Gonz\'alez E., Gallegos
J.E., Arg\"ueso F., Cay\'on L., Sanz J.L., 2002, MNRAS, 336, 22}
%
\bibitem{martinez-gonzalez_03}{Mart\'\i nez-Gonz\'alez E., Diego J.M., Vielva P. \& Silk J., 2003, MNRAS, 345, 1101}
%
\bibitem{martinez-gonzalez_vielva_05} Mart\'\i nez-Gonz\'alez E. \& Vielva P.,
2005: The Cosmic Microwave Background Anisotropies: Open Problems. In:
\textit{JENAM Workshop: The Many Scales in the Universe}, ed. by C. del
Toro et al., (Kluwer), astro-ph/0510003 
%
\bibitem{brain} Masi S. et al., 2005, EAS Publications Series,
14, 87-92
%
\bibitem{mather_94}{Mather J.C. et al. 1994, ApJ, 420, 439}
%
\bibitem{mather_99}{Mather J.C. et al. 1999, ApJ, 512, 511}
%
\bibitem{mcdonald_04} McDonald P. et al., 2005, ApJ, 635, 761,
astro-ph/0407377 
%
\bibitem{mcewen_05} {McEwen J. D., Hobson M. P., Lasenby A. N. \& Mortlock D. 
J., 2005, MNRAS, 359, 1583}
%
\bibitem{mcewen_06a} McEwen J.D., Vielva P., Hobson M.P., Mart\'\i
  nez-Gonz\'alez E. \& Lasenby A.N., 2006, MNRAS, submitted,
  astro-ph/0602398
%
\bibitem{mcewen_06b} {McEwen J. D., Hobson M. P., Lasenby A. N. \& Mortlock D. 
J., 2005, MNRAS, 359, 1583}
%
\bibitem{mennella_04}{Mennella A. et al., 2004, Recent Research
  Developments in Astronomy and Astrophysics, submitted, astro-ph/0402528}
%
\bibitem{mollerach_95} Mollerach S., Gangui A., Luchin F.,
Matarrese S., 1995, ApJ, 453, 1
%
\bibitem{mollerach_99}{Mollerach S., Mart\'{\i}nez V. J., Diego J. M.,
Mart\'{\i}nez-Gonz\'alez E., Sanz J. L., Paredes S., 1999, ApJ, 525, 17}
%
\bibitem{monteserin_05} Monteser\'{\i}n C., Barreiro R.B., Sanz,
  J.L. \& Mart\'\i nez-Gonz\'alez E., 2005, MNRAS, 360, 9
%
\bibitem{monteserin_06} Monteser\'{\i}n C., Barreiro R.B., Mart\'\i
  nez-Gonz\'alez E. \& Sanz J.L., 2006, MNRAS, 371, 312
%
\bibitem{montroy_05} Montroy T.E. et al., 2006, ApJ, 647, 813, astro-ph/0507514
%
\bibitem{mukherjee_00} Mukherjee P., Hobson M.P. \& Lasenby
A.N., 2000, MNRAS, 318, 1157-1163
%
\bibitem{mukherjee_wang_04} {Mukherjee P. \& Wang Y., 2004, ApJ, 613, 51}
%
\bibitem{naselsky_novikov_98} Naselsky P.D. \& Novikov D.I., 1998,
  ApJ, 507, 31 
%
\bibitem{naselsky_03} Naselsky P.D., Doroshkevich A.G. \& Verkhodanov
  O.V., 2003, ApJL, 599, L53 
%
\bibitem{naselsky_05} Naselsky P., Chiang L.-Y., Olesen P. \& Novikov
  I., 2005, Phys. Rev. D, 72, 063512
%
\bibitem{nolta_04} {Nolta M. R. et al., 2004, ApJ, 608, 10}
%
\bibitem{novikov_00} Novikov D., Schmalzing J. \& Mukhanov
V.F., 2000, A\& A, 364, 17-25 
%
\bibitem{OV} Ostriker J.P. \& Vishniac E.T., 1986, ApJ, 306, 510 
%
\bibitem{pando_98} Pando J., Valls-Gabaud D. \& Fang L.-Z., 1998,
Physical Review Letters, 81, 4568-4571 
%
\bibitem{park_01} Park C.-G., Park C., Ratra B. \& Tegmark
M., 2001, ApJ, 556, 852, astro-ph/0102406 
%
\bibitem{park_04}{Park C.-G., 2004, MNRAS, 349, 313}
%
\bibitem{patanchon_05} Patanchon G., Cardoso J.-F., Delabrouille J. \&
  Vielva P., 2005, MNRAS, 364, 1185
%
\bibitem{percival_01}{Percival W.J. et al., 2001, MNRAS, 327, 1297}
%
\bibitem{perlmutter_99} Perlmutter S., 1999, ApJ, 517, 565
%
\bibitem{phillips_kogut_01} Phillips N.G. \& Kogut A., 2001, ApJ,
548, 540-549. 
%
\bibitem{piacentini_05} Piacentini F. et al., 2005, ApJ, 647, 823,
  astro-ph/0507507 
%
\bibitem{planck} The Plank Consortia, 2005, in Planck: The Scientific
  Programme, European Space Agency, ESA-SCI(2005)1
%
\bibitem{pogosian_03} Pogosian, L., Tye, S.-H.H., Wasserman, I. \&
  Wyman, M., 2003, Phys. Rev. D, 68, 0235506 
%
\bibitem{polarbear} POLARBEAR, http://bolo.berkeley.edu/polarbear
%
\bibitem{polenta_02} Polenta G. et al., 2002, ApJ, 572, L27, astro-ph/0201133
%
\bibitem{quiet} QUIET, http://quiet.uchicago.edu
%
\bibitem{rapetti_05}{Rapetti D., Steven S.W. \& Weller J., 2005, MNRAS, 360, 555}
%
\bibitem{readhead_04a}{Readhead A.C.S. et al., 2004, ApJ, 609, 498}
%
\bibitem{readhead_04b} {Readhead A.C.S. et al., 2004, Science, 306, 836}
%
\bibitem{riazuelo_04} Riazuelo, A., Uzan, J.-P., Lehoucq, R. \& Weeks, J., 
2004, Phys. Rev. D, 69, 103514, astro-ph/0212223 
%
\bibitem{riess_98}{Riess et al., 1998, ApJ, 116, 1009}
%
\bibitem{riess_01}{Riess et al., 2001, ApJ, 560, 49}
%
\bibitem{rocha_01}{Rocha G., Magueijo J., Hobson M. \& Lasenby A., 2001,
Phys. Rev. D, 64, 063512, astro-ph/0008070}
%
\bibitem{rocha_04}{Rocha G., Cay\'on L., Bowen R., Canavezes A., Silk
J., Banday A.J. \& Gorski K.M., 2004, MNRAS, 351, 769, astro-ph/0205155}
%
\bibitem{roukema_00} Roukema, B., 2000, MNRAS, 312, 712
%
\bibitem{roukema_04}{Roukema B.F., Lew B., Cechowska M., Marecki A. \&
Bajtlik S., 2004, astro-ph/0402608}
%
\bibitem{RS} Rees M.J. \& Sciama D.W., 1968, Nature, 517, 611
%
\bibitem{rubino_06} Rubi\~no-Mart\'\i n et al., 2006, MNRAS, 369, 909
%
\bibitem{SW} {Sachs R. K. \& Wolfe A. M., 1967, ApJ, 147, 73}
%
\bibitem{santos_02} Santos M.G. et al., 2002, Phys. Rev. Lett., 88,
241302, astro-ph/0107588 
%
\bibitem{sanz_97}{Sanz J.L., 1997, in The Cosmic Microwave Background, eds. 
C.H. Lineweaver et al., Kluwer Academic Publishers, p. 33}
%
\bibitem{savage_04} Savage R. et al., 2004, MNRAS, 349, 973
%
\bibitem{scannapieco_99} Scannapieco, E., Levin and Silk, J. 1999, MNRAS, 
303, 797
%
\bibitem{schmalzing_gorski_98} Schmalzing J. \& Gorski K.M., 1998,
MNRAS, 297, 355.  
%
\bibitem{scott_94}{Scott D.H., Srednicki M. \& White M., 1994, ApJ, 421, 5}
%
\bibitem{shandarin_02} Shandarin S.F., Feldman H.A., Xu Y. \&
Tegmark M., 2002, ApJS, 141,1, astro-ph/0107136 
%
\bibitem{seljak_05}{Seljak U. et al., 2005, Phys. Rev. D, 71, 103515}
%
\bibitem{seljak_hirata_04} Seljak U. \& Hirata C.M., 2004,
Phys. Rev. D, 69, 043005
%
\bibitem{seljak_03} Seljak U., Sugiyama N., White M. \& Zaldarriaga
M., 2003, Phys. Rev. D, 68, 3507
%
\bibitem{CMBFAST} {Seljak U. \& Zaldarriaga M., 1996, ApJ, 469, 437 (http://www.cmbfast.org)}
%
\bibitem{seljak_zaldarriaga_99} Seljak U. \& Zaldarriaga M., 1999,
Phys. Rev. D60, 043504. 
%
\bibitem{silk_68}{Silk J., 1968, ApJ, 151, 459}
%
\bibitem{skrutskie_97} Skrutskie M.F. et al., 1997, in The Impact of
Large Scale Near-IR Sky Survey, eds. Gaz\'on F. et al., Kluwer,
Dordrecht, p. 187  
%
\bibitem{slosar_seljak_04} Slosar A. \& Seljak U., 2004, Phys. Rev. D,
  70, 083002
%
\bibitem{smith_04} Smith S. et al., 2004, MNRAS, 352, 887
%
\bibitem{smoot_92} {Smoot G. F. et al., 1992, ApJL, 396, L1}
%
\bibitem{WMAP:parameters} {Spergel D. N. et al., 2003, ApJS, 148, 175}
%
\bibitem{SZ}{Sunyaev R.A. \& Zeldovich Y.B., 1972, Comm. Astrophys. Space 
Phys., 4, 173}
%
\bibitem{schwartz_04} Schwartz D.J., Starkman, G.D., Huterer D. \&
Copi C.J., 2004, Phys. Rev. Lett., 93, 1301  
%
\bibitem{starobinsky_93} Starobinsky A.A., 1993, JETP Lett., 57, 622
%
\bibitem{tegmark_00}{Tegmark M., Eisenstein D.J., Hu W. \& de Oliveira-Costa 
A., 2000, ApJ, 530, 133}
%
\bibitem{tegmark_03} Tegmark M., de Oliveria-Costa A., Hamilton A.J.,
  2003, Phys. Rev. D, 68, 123523
%
\bibitem{tegmark_04a}{Tegmark M. et al., 2004, ApJ, 606, 702}
%
\bibitem{tegmark_04b}{Tegmark M. et al., 2004, Phys. Rev. D, 69, 103501}
%
\bibitem{tenorio_99}{Tenorio L., Jaffe A.H., Hanany S., Lineweaver
C.H., 1999, MNRAS, 310, 823}
%
\bibitem{tonry_03}{Tonry J.L. et al., 2003, ApJ, 594, 1}
%
\bibitem{tucci_05}{Tucci M., Mart\'\i nez-Gonz\'alez E., Vielva P. \& Delabrouille J., 
2005, MNRAS, 360, 935}
%
\bibitem{toffolatti_05}{Toffolatti L., Negrello M., Gonz\'alez-Nuevo J., de Zotti G., Silva L., Granato G.L. \& Arg\"ueso F., 2005, A\&A, 438, 475 
astro-ph/0410605}
%
\bibitem{tojeiro_06} Tojeiro R., Castro P.G., Heavens A.F. \& Gupta
  S., 2006, MNRAS, 365, 265 
%
\bibitem{tomita_05} Tomita K., 2005, Phys. Rev. D, 72, 103506
%
\bibitem{turok_spergel_90} Turok N. \& Spergel D., 1990, Phys. Rev. D,
  64, 2736
%
\bibitem{uzan_99} Uzan, J.-P., Lehoucq, R. \& J.-P. Luminet, J.-P. 1999. In 
\textit{Proceedings of the XIXth Texas Symposium}, Paris 14-18 December 1998, 
    Eds. E. Aubourg, T. Montmerle, J. Paul and P. Peter, article-no: 04/25
%
\bibitem{vale_05} Vale C., 2005, ApJL, submitted, astro-ph/0509039 
%
\bibitem{vielva_01}{Vielva P., Barreiro R. B., Hobson M. P., 
Mart\'\i nez-Gonz\'alez E., 
Lasenby A. N., Sanz J. L. \& Toffolatti L., 2001, MNRAS, 328, 1}
%
\bibitem{vielva_04} {Vielva P., Mart\'\i nez-Gonz\'alez E., Barreiro R.B., 
Sanz J.L. \& Cay\'on L., 2004, ApJ, 609, 22}
%
\bibitem{vielva_06} {Vielva P., Mart\'\i nez-Gonz\'alez E. \& Tucci M., 2006,
MNRAS, 365, 891, astro-ph/0408252}
%
\bibitem{vilenkin_shellard_94} Vilenkin A. \& Shellard E.P.S., 1994,
  \textit{Cosmic Strings and Other Topological Defects}, Cambridge
  University Press 
%
\bibitem{vishniac_87} Vishniac E.T., 1986, ApJ, 322, 597
%
\bibitem{wandelt_98}{Wandelt B.D., Hivon E. \& Gorski K.M., 1998:
Topological Analysis of High-Resolution CMB Maps. In:
\textit{Fundamental Parameters in Cosmology}, proceedings of the
XXXIIIrd Rencontres de Moriond, ed. by Tran Thanh Van
(Eds. Fronti\`eres 1998), astro-ph/9803317} 
%
\bibitem{watson_05}{Watson R.A., Rebolo R., Rubiño-Mart\'\i n J.A., Hildebrant
S., Guti\'errez C.M., Fern\'andez-Cerezo S., Hoyland R.J. \& Battistelli
E.S., 2005, ApJ, 624, L89, astro-ph/0503714}
%
\bibitem{weeks_98} Weeks, J. 1998, Class. Quant. Grav., 15, 2599
%
\bibitem{weeks_04} Weeks J.R., 2004, astro-ph/0412231 
%
\bibitem{white_05} White N.E., 2005, Adv. Space Res., 35, 96
%
\bibitem{wiaux_05}{Wiaux Y., Jaques L. \& Vandergheynst P., 2005, ApJ,
submitted, astro-ph/0508516}
%
\bibitem{wiaux_06}{Wiaux Y., Vielva P., Mart\'\i nez-Gonz\'alez E. \&
Vandergheynst P., 2006, Phys. Rev. Lett., 96, 151303}
%
\bibitem{wu_01a} Wu J.H.P.: New Method of Extracting non-Gaussian
Signals in the CMB. In: \textit{20th Texas Symposium on relativistic
astrophysics, Austin, Texas, 10-15 December, 2000}, vol. 586, ed. by
J.C. Wheeler and H. Martel (AIP Conference Proceedings), p. 211
%
\bibitem{wu_01b} Wu, J.H.P., Balbi, A., Borril, J., Ferreira, P.G.,
Hanany, S., Jaffe, A.H.,  
Lee, A.T., Rabii, B., Richards, P.L., Smoot, G.F., Stompor, R. \& Winant, C.D.
2001, Phys. Rev. Lett., 87, 251303, astro-ph/0104248.
%
\bibitem{wu_05} Wu, J.-H.P., 2005, astro-ph/0501239
%
\bibitem{wyman_05} Wyman M., Pogosian L. \& Wasserman I., 2005,
  Phys. Rev. D, 72, 023513
%
\bibitem{yoshida_01} Yoshida N., Sheth R. K. \& Diaferio A., 2001,
MNRAS, 328, 669, astro-ph/0104332  
%
\bibitem{zaldarriaga_seljak_97}{Zaldarriaga M. \& Seljak U., 1997, Phys. Rev. D, 55, 1830}
%
\bibitem{zaldarriaga_seljak_98}{Zaldarriaga M. \& Seljak U., 1998, Phys. Rev. D, 58, 023003}
%




\end{thebibliography}
\end{document}